\documentclass[11pt,a4paper]{article}
\usepackage{graphicx}
\usepackage{float}\vspace{.7cm}
\usepackage{afterpage}
\usepackage{epsfig,cite}
\usepackage{amssymb}
\usepackage{amsmath}
\usepackage{dsfont}
\usepackage{multirow}
\usepackage{url,hyperref}
\usepackage{booktabs}
\usepackage{tabularx}
\usepackage{xcolor}
\usepackage{bm}
\usepackage{textcomp}
\usepackage{caption}
\usepackage{slashed}
\usepackage{changes}
\usepackage{url}
\usepackage{hyperref}
\usepackage[symbol]{footmisc}
\usepackage{subfig}
\usepackage{braket}
\usepackage{lscape}
\usepackage{leftindex}
\usepackage{lipsum}
\usepackage{cancel}
\usepackage{standalone}
\usepackage{hhline}
\usepackage{diagbox}
\usepackage{rotating}

\newcommand{\e}{\varepsilon}
\newcommand{\ta}{\left(}
\newcommand{\qa}{\left[}

\newcommand{\tc}{\right)}
\newcommand{\qc}{\right]}

\renewcommand{\[}{\begin{equation}}
\renewcommand{\]}{\end{equation}}
\newcommand{\nb}{{\bar{n}}}

\usepackage[most]{tcolorbox}

\usepackage{dutchcal}
\usepackage{tikz}
\usetikzlibrary{decorations.pathmorphing}
\usetikzlibrary{decorations.markings}
\usetikzlibrary{arrows.meta,arrows}
\usetikzlibrary{calc}
\usetikzlibrary{math}
\usetikzlibrary{fpu}
\definecolor{myGreen}{rgb}{0.18,0.763,0.324}
\definecolor{myWhite}{rgb}{0.98,0.98,0.98}
\definecolor{myGray}{rgb}{0.7,0.7,0.75}
\definecolor{myGold}{rgb}{0.8,0.64,0.24}
\definecolor{myPurple}{rgb}{1,0.3,0.9}
\definecolor{myMauve}{rgb}{0.58,0,0.82}

\begin{document}

\begin{flushright}
IPARCOS-UCM-25-004\\
DESY-25-024
\end{flushright}

\begin{center} 
  {\Large \bf One-loop matching for leading-twist generalised\\
  \vspace{5pt}

  transverse-momentum-dependent distributions}
  \vspace{.7cm}

Valerio~Bertone$^1$\footnote{valerio.bertone@cea.fr}, Miguel G. Echevarria$^{2,3}$\footnote{miguel.garciae@ehu.eus}, Oscar del Rio$^4$\footnote{oscgar03@ucm.es}, Simone Rodini$^5$\footnote{simone.rodini@desy.de}

\vspace{.3cm}
{\it ~$^1$IRFU, CEA, Universit\'e Paris-Saclay, F-91191 Gif-sur-Yvette, France.}\\
{\it ~$^2$Department of Physics, University of the Basque Country UPV/EHU, 48080 Bilbao, Spain.}\\
{\it ~$^3$EHU Quantum Center, University of the Basque Country UPV/EHU.}\\
{\it ~$^4$Department of Theoretical Physics \& IPARCOS,\\Complutense University of Madrid, 28040 Madrid, Spain.}\\
{\it ~$^5$Deutsches Elektronen-Synchrotron DESY, Notkestr. 85, 22607 Hamburg, Germany.}
\end{center}

\begin{center}
  {\bf \large Abstract}\\
\end{center}

We present the one-loop matching coefficients necessary to match all of the leading-twist generalised transverse-momentum-dependent distributions (GTMDs) onto generalised parton distributions (GPDs). Matching functions are extracted by computing the first radiative corrections to partonic bilocal correlators with staple-like Wilson lines, as appropriate for high-energy collisions. These correlators are characterised by a transverse displacement and skewed kinematics of external states. Using the proton helicity basis, they are parametrised in terms of GTMDs, which are subsequently related to leading-twist GPDs. Our results provide new insights into the complex dynamics of GTMDs generated by radiative corrections. In particular, we show that time-reversal even and odd contributions to GTMDs in the so-called ERBL region mix both under matching and evolution. Finally, we present a selection of numerical results and comment on the quantitative behaviour of GTMDs.

\clearpage

\tableofcontents

\renewcommand*{\thefootnote}{\arabic{footnote}}
\setcounter{footnote}{0}

\section{Introduction}\label{sec_introduction}

A key ingredient in all modern phenomenological extractions of
transverse-momentum-dependent distributions (TMDs)
\cite{Bacchetta:2017gcc,Scimemi:2017etj,Scimemi:2019cmh,Bacchetta:2022awv,Cammarota:2020qcw,Echevarria:2020hpy,Bury:2022czx,Cerutti:2022lmb,Moos:2023yfa}
is a set of perturbative matching coefficients that encode the operator product
expansion of TMD operators in terms of collinear light-cone
operators~\cite{Collins:2011zzd,Moos:2020wvd}. On the theoretical side, this set
of relations must be satisfied by realistic models. On the phenomenological
side, these relations allow for simpler modeling and reduce parametric freedom
of input models, thus increasing quality and time requirements for extractions
of TMDs from experimental data.

In the case of forward TMDs, which are involved in inclusive processes, a wealth
of perturbative knowledge is currently available for the leading-twist matching
coefficients~\cite{Ebert:2020qef,Luo:2020epw,Gutierrez-Reyes:2018iod,
Gutierrez-Reyes:2019rug,Gutierrez-Reyes:2017glx,Buffing:2017mqm,
Bacchetta:2013pqa,Kanazawa:2015ajw,Scimemi:2018mmi,
Meissner:2007rx,Boer:2003cm,Ji:2006ub,Zhou:2009jm,Dai:2014ala,Scimemi:2019gge,Moos:2020wvd}
and some information for the next-to-leading-twist matching
coefficients~\cite{Rein:2022odl,Rodini:2022wki,Rodini:2023mnh}. In contrast, in
the case of non-forward TMDs, a.k.a. generalised TMDs (GTMDs), very little
perturbative information is currently available~\cite{Bertone:2022awq}. GTMDs
have been introduced and formalised in
Refs.~\cite{Meissner:2008ay,Meissner:2009ww,Lorce:2013pza,Kanazawa:2014nha,Echevarria:2016mrc}
and one of their most interesting aspects is their ability to give access to the
orbital angular momentum of
quarks~\cite{Lorce:2011kd,Hatta:2011ku,Hagiwara:2016kam,Hagiwara:2017fye}.
Although several proposals to probe GTMDs have been put forward in recent years (see, \textit{, e.g.,}, Refs.~\cite{Hatta:2016dxp, Hatta:2016aoc,
Hagiwara:2017ofm,Ji:2016jgn,Bhattacharya:2017bvs,
Bhattacharya:2018lgm,Boer:2021upt,Boer:2023mip,Bhattacharya:2023yvo,Bhattacharya:2023hbq})
and a proof of factorisation has recently been
obtained~\cite{Echevarria:2022ztg}, GTMDs remain elusive objects, very
challenging to access experimentally due to the exclusive nature of the
processes they are involved in. In order to perform any exploratory study of
their phenomenological impact, it is essential to have a solid theoretical
foundation, so that relevant details are not washed away by model freedom.

Our aim with this work is to present the complete set of one-loop, \textit{i.e.}
$\mathcal{O}(\alpha_s)$, corrections to the matching relations between
leading-twist GTMDs and generalised parton distributions (GPDs) by completing,
expanding, and to some extent amending the work of Ref.~\cite{Bertone:2022awq}.
Going beyond $\mathcal{O}(\alpha_s)$ is, at present, a superfluous step, which is
computationally very intensive due to the non-forward nature of GTMDs. Going
beyond leading-twist accuracy is similarly superfluous, as it would produce a set
of matching relations that involve a large number of completely unknown
twist-three GPDs. Therefore, for the foreseeable future, the only potentially
relevant matching coefficients are those presented in this work.

The work is organised as follows: in Sec.~\ref{sec_definitions_and_conventions},
we summarise all relevant definitions and conventions used in the paper; in
Sec.~\ref{sec_matching_op_level}, we present the results for the matching
coefficients and, in Sec.~\ref{sec_matching_param}, we construct the matching on
GPDs for all twist-two GTMDs in a given parametrisation. In
Sec.~\ref{sec_reconstructingGTMDs}, we provide the master formula for matching
and evolution, which we use in Sec.~\ref{sec_numerical_results} to present
numerical results. In Sec.~\ref{sec_conclusion}, we draw our conclusions.

\section{Definitions and conventions}
\label{sec_definitions_and_conventions}

We start by introducing two light-cone vectors, $n$ and $\nb$, defined in a way
that $n^2 = \nb^2 = 0 $ and $(\nb n) = 1$,\footnote{Scalar products between
four-vectors in Minkowski space are expressed as $a^\mu b_\mu \equiv (ab)$.} and
which define the longitudinal plane. The transverse plane orthogonal to $n$ and
$\nb$ is parametrised by two additional vectors, $R$ and $L$, which satisfy the
following orthonormality conditions:
\[
R^2 = L^2 = (R\,n) = (L\,n) = (R\,\nb)= (L\,\nb) = 0, \quad (R\,L) = -1.
\]
With this basis of vectors at hand, any generic four-vector $v$ can be
decomposed as $v=v^+\overline{n}+v^-n-v_R L - v_L R$, where $v^+ = (vn)$, $v^- =
(v\nb)$, $v_R = (vR)$ and $v_L = (vL)$. An explicit implementation of the basis
vectors is given by:
\[
\begin{array}{l}
\displaystyle n^\mu=\frac{1}{\sqrt{2}} (1,0,0,-1),\quad \nb^\mu=\frac{1}{\sqrt{2}}(1,0,0,1),\\
\\
\displaystyle R^\mu = -\frac{1}{\sqrt{2}}(0,1,i,0), \quad L^\mu = -\frac{1}{\sqrt{2}}(0,1,-i,0)\,.
\end{array}
\]
The scalar product between the generic vector $v$ and a purely transverse vector
$w_T$, \textit{i.e.} a four-vector characterised by $w_T^+=w_T^-=0$, produces
$(vw_{T})=-\bm{v}\cdot \bm{w}_T$, where the r.h.s. denotes the scalar product
between two two-dimensional vectors in Euclidean space. Moreover, given any
generic tensor $T^{\mu_1\mu_2\dots}$, we will use the following shorthand
notation for its contraction with the four-vectors $a_{\mu_1}$, $b_{\mu_2}$,
\dots: $a_{\mu_1} b_{\mu_2}\dots T^{\mu_1\mu_2\dots} \equiv T^{ab\dots}$.

We now turn to the definition of the partonic correlators that we will use to
extract the matching coefficients. The relevant quark and gluon GTMD correlators
read:
\begin{align}
  \label{QuarkCorrelator}&\mathcal{F}^{[Y]}_{q/H}(x,\xi,\bm{b},\bm{\Delta}_T) \\
  & \notag \qquad = S_q^{-\frac12}(\bm{b})\int \frac{dz}{2\pi} e^{-ixp^+z} \left\langle p + \frac{\Delta}{2},\lambda'\right|\bar{\psi}_i\ta\frac{zn}{2}+\frac{\bm{b}}{2}\tc Y_{ij}\mathcal{W}_s\psi_j\ta-\frac{zn}{2}-\frac{\bm{b}}{2}\tc\left|p - \frac{\Delta}{2},\lambda\right\rangle\,,\\
  \label{GluonCorrelator} & \mathcal{F}^{[Y]}_{g/H}(x,\xi,\bm{b},\bm{\Delta}_T) \\
& \notag \qquad = S_g^{-\frac12}(\bm{b})\int \frac{dz}{2\pi} \frac{e^{-ixp^+z}}{xp^+} \left\langle p + \frac{\Delta}{2},\lambda'\right|F^{\mu n}\ta\frac{zn}{2}+\frac{\bm{b}}{2}\tc Y_{\mu\nu}\mathcal{W}_s F^{\nu n}\ta-\frac{zn}{2}-\frac{\bm{b}}{2}\tc\left|p - \frac{\Delta}{2},\lambda\right\rangle\,,
\end{align}
where $p-\Delta/2$ and $\lambda$ ($p+\Delta/2$ and $\lambda'$) denote momentum
and helicity of the incoming (outgoing) hadron $H$, respectively. As usual, $x$
is the fraction of momentum $p$ along the longitudinal direction carried by the
parton ($q$ or $g$); while $\xi = -(\Delta n) / 2(pn)$, a.k.a. skewness
parameter, is the fraction of the momentum transfer $\Delta$  w.r.t. $p$ along
the longitudinal direction. $\psi_j$ is the $j$-th component of the Dirac spinor
for the quark $q$, and $F^{\mu\nu}=t_a F_a^{\mu\nu}$, with $F_a^{\mu\nu}$ the
gluon field strength for colour configuration $a$ and $t_a$ the corresponding
SU(3) group generator. In both correlators, the bilocal operator is
characterised by the transverse displacement $\bm{b}$. Therefore, the variables
$x$, $\xi$, $\bm{b}$, and $\bm{\Delta}_T$, \textit{i.e.} the transverse
component of $\Delta$, fully define the kinematics of the correlator.
$\mathcal{W}_s$ is the staple-like Wilson line which connects the two space-time
points of the bilocal operator and which runs along the light-cone direction
defined by $n$ up to $s\infty$, where the sign $s=\pm1$ depends on the process
under consideration. The colour representation of the Wilson line is the
fundamental one for the quark correlator and the adjoint one for the gluon
correlator. Important ingredients of the definitions in
Eqs.~(\ref{QuarkCorrelator}) and~(\ref{GluonCorrelator}) are the soft functions
$S_i$, with $i=q,g$, whose presence is necessary to ensure the cancellation of
the so-called rapidity divergences generated by the partonic
correlators~\cite{Echevarria:2016mrc,Bertone:2022awq}. $S_i$ is a universal
perturbative quantity currently known up to three-loops in perturbative
QCD~\cite{Li:2016ctv}. In this work, we will only need the one-loop result.

The corresponding operator definitions for quark and gluon GPD correlators are:
\begin{align}
\label{QuarkGPDCorrelator}
& F^{[\Gamma]}_{q/H}(x,\xi,\bm{\Delta}_T) = \int \frac{dz}{2\pi} e^{-ixp^+z} \left\langle p + \frac{\Delta}{2},\lambda'\right|\bar{\psi}_i\ta\frac{zn}{2}\tc \Gamma_{ij}\mathcal{W}\psi_j\ta-\frac{zn}{2}\tc\left|p - \frac{\Delta}{2},\lambda\right\rangle\,,\\
\label{GluonGPDCorrelator}
  & F^{[\Gamma]}_{g/H}(x,\xi,\bm{\Delta}_T) = \int \frac{dz}{2\pi} \frac{e^{-ixp^+z}}{xp^+} \left\langle p + \frac{\Delta}{2},\lambda'\right|F^{\mu+}\ta\frac{zn}{2}\tc \Gamma_{\mu\nu}\mathcal{W} F^{\nu+}\ta-\frac{zn}{2}\tc\left|p - \frac{\Delta}{2},\lambda\right\rangle\,,
\end{align}
which are obtained from the GTMD correlators in Eqs.~(\ref{QuarkCorrelator})
and~(\ref{GluonCorrelator}) by setting $\bm{b}=\mathcal{0}$ (which also implies
that the soft functions $S_i$ reduce to unity). Because of the vanishing
transverse displacement, the Wilson line $\mathcal{W}$ no longer depends on the
light-cone direction, hence we dropped the subscript $s$.

Finally, we define the projectors $Y$ and $\Gamma$ that enter into the
definitions of the GTMD and GPD correlators, respectively, and which run over
the same set of Dirac (for quarks) and Lorentz (for gluons) structures. They are
given by:\footnote{Projectors obtained through the replacement $R\to L$ project
out the same GTMD/GPD correlators, hence need not be considered separately.}
\begin{align}
\label{QuarkProjectors}\text{quarks}: \quad &Y_{ij},\Gamma_{ij}\in\left\{\frac{\gamma^+}{2}, \frac{\gamma^+\gamma_5}{2}, \frac{i\sigma^{R+}\gamma_5}{\sqrt{2}}\right\}_{ij}\,,\\
\label{GluonProjectors}\text{gluons}: \quad &Y^{\mu\nu},\Gamma^{\mu\nu}\in\left\{-g_T^{\mu\nu}, -i\e^{\mu\nu}_T, 2R^\mu R^\nu \right\}\,,
\end{align}
where:
\begin{align}
  g_{T}^{\mu\nu} & = g^{\mu\nu} - n^\mu \nb^\nu - n^\nu \nb^\mu = R^\mu L^\nu + R^\nu L^\mu\,,\\
  \e_T^{\mu\nu} & = \e^{\mu\nu \nb n}=R^\mu L^\nu - R^\nu L^\mu\,,
\end{align}
with $\e_T^{12} = +1$. The projectors in Eqs.~(\ref{QuarkProjectors})
and~(\ref{GluonProjectors}) are engineered to project out the
\textit{leading-twist} (twist-two) parts of the correlators. Specifically, we
have that: $\gamma^+/2$ and $-g_T^{\mu\nu}$ project out the unpolarised ($U$)
components, $\gamma^+\gamma_5/2$ and $-i\e_T^{\mu\nu}$ project out the
longitudinally-polarised ($L$) components, and $i\sigma^{R+}\gamma_5/\sqrt{2}$
and $2R^\mu R^\nu$ project out the transversely(quark)/linearly(gluon) polarised
($T$) components. Note that we use the letter $L$ both as a label for
longitudinal polarisations and as one of the two basis vectors for the
transverse space. The meaning of each instance of $L$ should however be clear
from context.

\section{Matching functions}
\label{sec_matching_op_level}

According to the operator-product expansion (OPE)
formalism~\cite{Collins:2011zzd,Scimemi:2018mmi,Moos:2020wvd}, the GTMD
correlators in Eqs.~(\ref{QuarkCorrelator}) and~(\ref{GluonCorrelator}) at small
$|\bm{b}|$ can be expanded in terms of correlators characterised by light-like
separations which can be identified with the GPD correlators in
Eqs.~(\ref{QuarkGPDCorrelator}) and~(\ref{GluonGPDCorrelator}). The OPE leading
term for GTMD correlators can thus be factorised as:
\begin{equation}
\begin{array}{rcl}
  \displaystyle \mathcal{F}^{[Y]}_{i/H}(x,\xi,\bm{b},\bm{\Delta}_T;\mu,\zeta)&=&\displaystyle \int_x^\infty\frac{dy}{y}\mathcal{C}^{Y/\Gamma}_{i/j}\left(y,\kappa,\bm{b};\mu,\zeta\right)F^{[\Gamma]}_{j/H}\left(\frac{x}{y},\xi,\bm{\Delta}_T;\mu\right)\\
  \\
    &\equiv& \displaystyle \left[\mathcal{C}^{Y/\Gamma}_{i/j}\otimes F^{[\Gamma]}_{j/H}\right]\left(x,\xi,\bm{b},\bm{\Delta}_T;\mu,\zeta\right)\,,
\end{array}
\label{MatchingOPE}
\end{equation}
where repeated flavour ($j$) and polarisation ($\Gamma$) indices are implicitly
summed over. An important feature of Eq.~(\ref{MatchingOPE}) is that the
matching functions $\mathcal{C}^{Y/\Gamma}_{i/j}$ are generally not diagonal in
polarisation space. This implies that, for any given GTMD polarisation $Y$, GPDs
with different polarisations $\Gamma$ can contribute to it through matching.

We notice that the matching functions $\mathcal{C}^{Y/\Gamma}_{i/j}$ can only
depend over $\xi$ through the ratio:
\[
\kappa = \frac{\xi}{x}.
\]
Importantly, by separating the correlators into their even and odd contributions under $x\to -x$, it
is always possible to reduce ourselves to considering $\kappa>0$.\footnote{This
is standard practice in the case of forward distributions, where one trades
quark distributions at negative values of $x$ for the corresponding anti-quark
distributions. Gluon distributions are instead either symmetric or
anti-symmetric under $x\to -x$, depending on the polarisation.} We also
introduced the scales $\mu$ and $\zeta$ which emerge from the renormalisation of
the UV and rapidity divergences of the correlators, respectively. For the exact
definition of the rapidity scale $\zeta$ we refer to App. \ref{AppendixPole}.

The functions $\mathcal{C}^{Y/\Gamma}_{i/j}$ admit the following power-series
expansion in terms of the strong-coupling constant $\alpha_s$:
\begin{equation}
\mathcal{C}^{Y/\Gamma}_{i/j}\left(y,\kappa,\bm{b};\mu,\zeta\right)=\delta_{ij}\delta_{Y\Gamma}\delta(1-y)+\sum_{n=1}^{\infty}\left(\frac{\alpha_s (\mu)}{4\pi}\right)^n \mathcal{C}^{Y/\Gamma,[n]}_{i/j}\ta y,\kappa,\bm{b};\mu,\zeta\tc\,.
\end{equation}
In the following, we will compute the one-loop contributions to this expansion,
\textit{i.e.} $\mathcal{C}^{Y/\Gamma,[1]}_{i/j}$, which can be generally
written as:
\begin{equation}\label{DecompositionMatchingCoefficient}
  \begin{array}{rcl}
\displaystyle \mathcal{C}^{Y/\Gamma,[1]}_{i/j}\ta y,\kappa,\bm{b};\mu,\zeta\tc &=& \displaystyle R^{Y/\Gamma,[1]}_{i/j}(y, \kappa)-\delta_{Y\Gamma}\mathcal{P}_{i/j}^{\Gamma,[0]}\ta y,\kappa\tc
    \log\left(\frac {\mu^2\bm{b}^2}{b_0^2}\right)\\
    \\
& -&\displaystyle 2\,\delta_{ij}\,\delta_{Y\Gamma}\,\delta(1-y)\,C_i\Bigg[
\frac{1}{2}\log^2\left(\frac{\mu^2\bm{b}^2}{b_0^2}\right) +\frac{\pi^2}{12} \\
\\
& -&\displaystyle \left(K_i +\log\left(\frac {\mu^2}{|1-\kappa^2|\zeta}\right) - is\pi\theta(\kappa-1)\right)\log\left(\frac{\mu^2 \bm{b}^2}{b_0^2}\right)
\Bigg]\,,
  \end{array}
\end{equation}
where $b_0=2e^{-\gamma_{\rm E}}$, with $\gamma_{\rm E}$ the Euler constant. The
imaginary part in the third line and the $\log|1-\kappa^2|$ emerges from the UV
renormalization of the transverse momentum operator, we refer to
App.~\ref{AppendixPole} for the details. The logarithm $\log|1-\kappa^2|$ diverges
when $x=\xi$. Using as reference the pion-nucleon double Drell-Yan discussed in
Ref.~\cite{Echevarria:2022ztg}, this kinematic point is outside the
factorisation region. Indeed, when $x=\xi$, one of the two partons from the
nucleon carries vanishing longitudinal momentum, leading to $Q^2\sim q_T^2$ for
the virtual boson momentum, which contradicts the factorisation assumption of
$Q^2\gg q_T^2$. This statement is valid to all orders in perturbation theory. Therefore, we conclude that, at least for the double Drell-Yan
process, one cannot probe the region $x\sim \xi$ within the transverse-momentum
factorisation. It is conceivable that a different factorization theorem (compared to TMD factorisation) could allow for the resummation of the large logarithms near $x=\xi$. However, its investigation goes beyond the scope of this work.

Coming back to Eq. \eqref{DecompositionMatchingCoefficient}, the functions
$\mathcal{P}_{i/j}^{\Gamma,[0]}$ are the one-loop GPD splitting functions (see
Refs.~\cite{Ji:1998pc,Diehl:2003ny,Kirch:2005tt,Diehl:2007zu,Bertone:2023jeh})
and the constants $K_i$ and $C_i$, for $i=q,g$ read:
\begin{equation}
  \begin{array}{ll}
    \displaystyle K_q = \frac{3}{2}\,,&\quad \displaystyle C_q = C_F,\\
    \\
    \displaystyle K_g = \frac{11}{6} - \frac{2n_f}{3} \frac{T_R}{C_A}\,, & \displaystyle \quad C_g = C_A\,.
  \end{array}
\end{equation}
The one-loop \textit{residual} functions $R^{Y/\Gamma,[1]}_{i/j}$ in
Eq.~(\ref{DecompositionMatchingCoefficient}) are computed here for the first
time for the full set of leading-twist
polarisations.\footnote{Ref.~\cite{Bertone:2022awq} presented results for
$\mathcal{C}^{U/U,[1]}_{i/j}$ only.} These functions can be further decomposed
as follows:
\begin{equation}
    R_{i/j}^{Y/\Gamma,[1]}(y, \kappa)=\theta(1-y)\left[\theta(1+\kappa)r_{i/j}^{Y/\Gamma,[1]}(y, \kappa)+s_{Y/\Gamma}\,\theta(1-\kappa)r_{i/j}^{Y/\Gamma,[1]}(y,-\kappa)\right],
    \label{eq:Rdecomposition}
\end{equation}
where $s_{Y/\Gamma}=-1$ when \textit{either} $Y$ \textit{or} $\Gamma$ is equal
to $L$ (so that $s_{L/L}=1$), while $s_{Y/\Gamma}=1$ otherwise. Results for all
of the functions $r_{i/j}^{Y/\Gamma,[1]}(y,\kappa)$ are collected in
Tab.~\ref{tab_rij}, where rows correspond to GTMD parton species and
polarisations $(i,Y)$, and columns to GPD parton species and polarisations
$(j,\Gamma)$. As anticipated above, this table clearly shows that GTMD and GPD
polarisations mix upon matching. For details on the computational methods used
to obtain the functions in Tab.~\ref{tab_rij}, we refer the reader to
Refs.~\cite{Bertone:2022frx,Bertone:2022awq,Bertone:2023jeh}.

\renewcommand{\arraystretch}{1.2}
\begin{sidewaystable}
    \centering
    \begin{tabular}{c||c|c|c|c|c|c}
      \diagbox{$i,Y$}{$j,\Gamma$}& $q,U$ & $g,U$ &  $q,L$ & $g,L$ & $q,T$ & $g,T$ \\
     \hhline{=#=|=|=|=|=|=}
       $q,U$     & $\displaystyle C_F\frac{1+\kappa}{\kappa (1+\kappa y)}$ & $\displaystyle \frac{y(\kappa+1)}{\kappa (1+\kappa y)^2 \left(1-\kappa y+is\delta\right)}$ & 0 & 0 & 0 & $\displaystyle \frac{y(1+\kappa)}{\kappa (1+\kappa y)^2 \left(1-\kappa y + is\delta\right)}$ \\
       $g,U$     & $\displaystyle - C_F\frac{1-\kappa^2}{\kappa (1+\kappa y)}$ & $\displaystyle C_A\qa\frac{2 (1+\kappa)}{(1+\kappa y)^2}-\frac{(1+\kappa)^2}{\kappa (1+\kappa y)(1-\kappa y +is\delta)}\qc$ & 0 & 0 & 0 & $\displaystyle -C_A\frac{2 y (1+\kappa) \left(1+\kappa^2\right)}{\kappa (1+\kappa y)^2(1-\kappa y +is\delta)}$ \\
       $q,L$     & 0 & 0 & $\displaystyle C_F \frac{1+\kappa}{\kappa (1+\kappa y)}$ & $\displaystyle \frac{1+\kappa}{\kappa (1+\kappa y)^2(1-\kappa y +is\delta)}$ & 0 & $\displaystyle \frac{y(1+\kappa)}{(1+\kappa y)^2(1-\kappa y +is\delta)}$ \\
       $g,L$     & 0 & 0 & $\displaystyle -\frac{2(\kappa+1)C_F}{\kappa (1+\kappa y)}$ & $\displaystyle - C_A\frac{4 (1+\kappa)}{\kappa (1+\kappa y)^2(1-\kappa y +is\delta) }$ & 0 & $\displaystyle -C_A\frac{4 y(1+\kappa) }{(1+\kappa y)^2(1-\kappa y +is\delta) }$ \\
       $q,T$     & 0 & 0 & 0 & 0 & 0 & 0 \\
       $g,T$     & $\displaystyle -C_F\frac{2 (1+\kappa)}{\kappa y (1+\kappa y)}$ & $\displaystyle -C_A\frac{2  (1+\kappa) \left(1+\kappa^2 y^2\right)}{\kappa y(1+\kappa y)^2(1-\kappa y +is\delta)}$ & $\displaystyle \frac{2 (\kappa+1)C_F }{1+\kappa y}$ & $\displaystyle \frac{4 (\kappa+1)C_A}{(1+\kappa y)^2(1-\kappa y +is\delta)}$ & 0 & $\displaystyle -C_A\frac{4 \kappa  y(1+\kappa)}{(1+\kappa y)^2(1-\kappa y +is\delta)}$
    \end{tabular}
    
    \caption{Results for $r^{Y/\Gamma}_{i/j}(y,\kappa)$. Rows correspond to the
    GTMD parton species and polarisation $(i,Y)$, while columns to the GPD
    parton species and polarisation $(j,\Gamma)$. The presence of the term
    $is\delta$ appearing in some denominators is discussed in
    App.~\ref{AppendixPole}.}
    \label{tab_rij}
\end{sidewaystable}

Given the decomposition in Eq.~(\ref{eq:Rdecomposition}), it is convenient to construct specific combinations of GTMDs that are more suited for a numerical implementation of the matching. We define the flavour non-singlet GTMD combination as:
\begin{equation}\label{eq:totalvalence}
  \mathcal{F}^{[Y],-} = \sum_{q=1}^{n_f}\left[\mathcal{F}_{q\leftarrow H}^{[Y]}-\mathcal{F}_{\bar{q}\leftarrow H}^{[Y]}\right]\,,
\end{equation}
\renewcommand{\arraystretch}{1.5}
and the flavour singlet GTMD combination as:
\begin{equation}
  \mathcal{F}^{[Y],+}=
\begin{pmatrix}
  \displaystyle \sum_{q=1}^{n_f}\left[\mathcal{F}_{q\leftarrow H}^{[Y]}+\mathcal{F}_{\bar{q}\leftarrow H}^{[Y]}\right]\\
  \mathcal{F}_{g\leftarrow H}^{[Y]}
\end{pmatrix}\,,
\end{equation}
where the index $q$ runs over the $n_f$ active quark flavors. Using these
combinations, the non-singlet residual function $R^{Y/\Gamma,-,[1]}$ acting on
the combination $\mathcal{F}^{[Y],-}$ takes the form:
\begin{equation}
 R^{Y/\Gamma,-,[1]}\left(y,\kappa\right) = \theta(1-y)
 R_1^{Y/\Gamma,-,[1]}\left(y,\kappa\right)+\theta(\kappa-1)
 R_2^{Y/\Gamma,-,[1]}\left(y,\kappa\right),
\label{eq:MatchingDecompositionforNonSinglet}
\end{equation}
where:
\begin{equation}
\begin{array}{rcl}
  \displaystyle R_1^{Y/\Gamma,-,[1]}\left(y,\kappa\right)&=&\displaystyle r_{q/q}^{Y/\Gamma}\left(y,\kappa\right)+r_{q/q}^{Y/\Gamma}\left(y,-\kappa\right)\,,\\
R_2^{Y/\Gamma,-,[1]}\left(y,\kappa\right)&=&\displaystyle -r_{q/q}^{Y/\Gamma}\left(y,-\kappa\right)+s_{\Gamma} r_{q/q}^{Y/\Gamma}\left(-y,-\kappa\right)\,,
\label{eq:R1andR2NS}
\end{array}
\end{equation}
with $s_\Gamma=1$ for $\Gamma=U,T$ and $s_\Gamma=-1$ for $\Gamma=L$. The singlet
residual function $R^{Y/\Gamma,+,[1]}$ acting on the combination
$\mathcal{F}^{[Y],+}$ is instead a $2\times2$ matrix in flavour space and its
components can be written as:
\begin{align}
\label{eq:MatchingDecompositionforSinglet}
&R_{i/j}^{Y/\Gamma,+,[1]}\left(y,\kappa\right) = 
\theta(1-y)\theta(1-\kappa)
R_{1,i/j}^{Y/\Gamma,+,[1]}\left(y,\kappa\right)+
\theta(y-1) \theta(\kappa-1) R_{2,i/j}^{Y/\Gamma,+,[1]}\left(y,\kappa\right)\\
& \notag +
\theta(1-y)\theta(\kappa-1)\ta \widehat{R}_{i/j}^{Y/\Gamma,+,[1]}\left(y,\kappa\right)+ \Sigma_{i/j}^{Y/\Gamma,+,[1]}(\kappa)\qa \text{PV}\ta\frac{1}{1-\kappa y}\tc-\frac{i\pi s}{\kappa} \delta\ta y-\frac{1}{\kappa}\tc \qc \tc,
\end{align}
where PV indicates the principal-value prescription and, as above, $s=\pm 1$ is
the light-cone direction of the Wilson line. The functions
$R_{1,i/j}^{Y/\Gamma,+,[1]}$ and $R_{2,i/j}^{Y/\Gamma,+,[1]}$ are defined as:
\begin{equation}
\begin{array}{rcl}
\displaystyle R_{1,i/j}^{Y/\Gamma,+,[1]}\left(y,\kappa\right)&=&\displaystyle r_{i/j}^{Y/\Gamma}\left(y,\kappa\right)+s_{Y/\Gamma}r_{i/j}^{Y/\Gamma}\left(y,-\kappa\right)\,,\\
\displaystyle R_{2,i/j}^{Y/\Gamma,+,[1]}\left(y,\kappa\right)&=&\displaystyle -s_{Y/\Gamma}r_{i/j}^{Y/\Gamma}\left(y,-\kappa\right)-s_{\Gamma} r_{i/j}^{Y/\Gamma}\left(-y,-\kappa\right)\,,
\label{eq:R1andR2SG}
\end{array}
\end{equation}
while the functions $\widehat{R}_{i/j}^{Y/\Gamma,+,[1]}$ and the coefficients
$\Sigma_{i/j}^{Y/\Gamma,+,[1]}$ derive from the equality:
\begin{equation}
  R_{1,i/j}^{Y/\Gamma,+,[1]}\left(y,\kappa\right)+R_{2,i/j}^{Y/\Gamma,+,[1]}\left(y,\kappa\right)=\widehat{R}_{i/j}^{Y/\Gamma,+,[1]}\left(y,\kappa\right)+\frac{\Sigma_{i/j}^{Y/\Gamma,+,[1]}\left(\kappa\right)}{1-\kappa y+is\delta}\,.
  \label{eq:PoleIsolation}
\end{equation}
This equality is meant to isolate the singularity at $y=1/\kappa$ exhibited by
the combination $R_{1,i/j}^{Y/\Gamma,+,[1]}+R_{2,i/j}^{Y/\Gamma,+,[1]}$ in the
term proportional to $\Sigma_{i/j}^{Y/\Gamma,+,[1]}$, while the term
$\widehat{R}_{i/j}^{Y/\Gamma,+,[1]}$ is regular at this point. We point out that
Eq.~\eqref{eq:MatchingDecompositionforSinglet} is engineered to avoid
overlapping regions between different terms in a way to make explicit the
absence of non-integrable singularities. We refer the reader to
App.~\ref{AppendixPole} for more details on the origin of the imaginary part in
the second line of Eq.~\eqref{eq:MatchingDecompositionforSinglet}. In view of an
implementation, a numerically amenable formulation of PV prescription, when
convoluted with a regular function $f$, is achieved as follows:
\[
\begin{array}{l}
\displaystyle \int_x^\infty dy\,\theta(1-y)\,\text{PV}\left(\frac{1}{1-\kappa y}\right)f(y) = \\
\\
\displaystyle \quad\int_x^1\frac{dy}{1-\kappa y}\left[f(y)-f\left(\frac{1}{\kappa}\right)\left(1+\theta(\kappa y-1)\frac{1-\kappa y}{\kappa y}\right)\right]
+f\left(\frac{1}{\kappa}\right)\frac{1}{\kappa}\log\left(\frac{\kappa(1-\kappa x)}{\kappa-1}\right)\,.
\end{array}
\]

We finally give explicit expressions for the terms involved in the r.h.s. of
Eqs.~(\ref{eq:MatchingDecompositionforNonSinglet})
and~(\ref{eq:MatchingDecompositionforSinglet}). As evident from
Tab.~\ref{tab_rij}, in the non-singlet case,
Eq.~(\ref{eq:MatchingDecompositionforNonSinglet}), there are only two
polarisation channels that get a $q/q$ contribution necessary to construct
$R_1^{Y/\Gamma,-,[1]}$ and $R_2^{Y/\Gamma,-,[1]}$: $U/U$ and $L/L$. Using
Eq.~(\ref{eq:R1andR2NS}), the corresponding expressions read: 
\begin{alignat}{3}
  &\begin{cases}
  R_1^{U/U,-,[1]}\left(y,\kappa\right) = \displaystyle C_F\frac{2 (1-y) }{1-\kappa^2 y^2}\,,\\
  \\
  R_2^{U/U,-,[1]}\left(y,\kappa\right) = \displaystyle C_F\frac{2y (1-\kappa)  C_F}{1-\kappa^2 y^2}\,,
  \end{cases}
  &&\begin{cases}
  R_1^{L/L,-,[1]}\left(y,\kappa\right) = \displaystyle C_F\frac{2 (1-y) }{1-\kappa^2 y^2}\,,\\
  \\
  R_2^{L/L,-,[1]}\left(y,\kappa\right) = \displaystyle C_F\frac{2 (1-\kappa) }{\kappa(1-\kappa^2 y^2)}\,.
  \end{cases}
\end{alignat}

In the singlet case, Eq.~(\ref{eq:MatchingDecompositionforSinglet}), explicit
expressions for $R_{1,i/j}^{Y/\Gamma,+,[1]}$, $R_{2,i/j}^{Y/\Gamma,+,[1]}$,
$\widehat{R}_{i/j}^{Y/\Gamma,+,[1]}$, and $\Sigma_{i/j}^{Y/\Gamma,+,[1]}$,
obtained by means of Eqs.~(\ref{eq:R1andR2SG}) and~(\ref{eq:PoleIsolation}), are
collected in Tabs.~\ref{tab_R1_definitive}, \ref{tab_R2_definitive},
\ref{tab_Rhat_definitive}, and \ref{tab_Sigma}, respectively.
\begin{table}[!ht]
  \centering\resizebox{\textwidth}{!}{
    \begin{tabular}{c||c|c|c|c|c|c}
      \diagbox{$i,Y$}{$j,\Gamma$}& $q,U$ & $g,U$ & $q,L$ & $g,L$ & $q,T$ & $g,T$ \\
     \hhline{=#=|=|=|=|=|=}
       $q,U$     & $C_F\frac{2 (1-y) }{1-\kappa^2 y^2}$ & $n_f\frac{4 y(1-y)}{\left(1-\kappa^2 y^2\right)^2}$ & $0$ & $0$ & $0$ & $n_f\frac{4 y(1-y)}{\left(1-\kappa^2 y^2\right)^2}$ \\
       $g,U$     & $C_F\frac{2 y \left(1-\kappa^2\right)  }{1-\kappa^2 y^2}$ & $-C_A\frac{8 \kappa^2 y(1-y) }{\left(1-\kappa^2 y^2\right)^2}$ & $0$ & $0$ & $0$ & $-C_A\frac{4 \left(1+\kappa^2\right) y(1-y) }{\left(1-\kappa^2 y^2\right)^2}$ \\
       $q,L$     & $0$ & $0$ & $C_F\frac{2 (1-y) }{1-\kappa^2 y^2}$ & $n_f\frac{4(1-y) }{\left(1-\kappa^2 y^2\right)^2}$ & $0$ & $n_f\frac{4 \kappa y(1-y)}{\left(1-\kappa^2 y^2\right)^2}$ \\
       $g,L$     & $0$ & $0$ & $-C_F\frac{4 (1-y) }{1-\kappa^2 y^2}$ & $-C_A\frac{8 (1-y) }{\left(1-\kappa^2 y^2\right)^2}$ & $0$ & $-C_A\frac{8 \kappa y(1-y) }{\left(1-\kappa^2 y^2\right)^2}$ \\
       $q,T$     & $0$ & $0$ & $0$ & $0$ & $0$ & $0$ \\
       $g,T$     & $-C_F\frac{4 (1-y) }{y \left(1-\kappa^2 y^2\right)}$ & $-C_A \frac{4 (1-y) \left(1+\kappa^2 y^2\right)}{y \left(1-\kappa^2 y^2\right)^2}$ & $C_F\frac{4 \kappa (1-y) }{1-\kappa^2 y^2}$ & $C_A\frac{8 \kappa (1-y) }{\left(1-\kappa^2 y^2\right)^2}$ & $0$ & $-C_A\frac{8 \kappa^2 y(1-y) }{\left(1-\kappa^2 y^2\right)^2}$ 
    \end{tabular}}

    \caption{Results for the functions $R^{Y/\Gamma,+,[1]}_{1,i/j}$ in the
    r.h.s. of Eq.~(\ref{eq:MatchingDecompositionforSinglet}). Rows correspond to
    the GTMD parton species and polarisation $(i,Y)$, while columns to the GPD
    parton species and polarisation $(j,\Gamma)$.}
    \label{tab_R1_definitive}
\end{table}

\begin{table}[!ht]
  \centering\resizebox{\textwidth}{!}{
    \begin{tabular}{c||c|c|c|c|c|c}
      \diagbox{$i,Y$}{$j,\Gamma$}& $q,U$ & $g,U$ & $q,L$ & $g,L$ & $q,T$ & $g,T$ \\
     \hhline{=#=|=|=|=|=|=}
       $q,U$     & $C_F\frac{2 (1-\kappa) }{\kappa(1-\kappa^2 y^2)}$ & $n_f\frac{4 y^2(1-\kappa) }{\left(1-\kappa^2 y^2\right)^2}$ & $0$ & $0$ & $0$ & $ n_f\frac{4y^2(1-\kappa) }{\left(1-\kappa^2 y^2\right)^2}$ \\
       $g,U$     & $-C_F\frac{2 \left(1-\kappa^2\right)}{\kappa \left(1-\kappa^2 y^2\right)}$ & $-2C_A\frac{1-\kappa}{1-\kappa^2y^2}\qa \frac{1-\kappa}{\kappa} +\frac{2(1+\kappa^2y^2)}{1-\kappa^2y^2}\qc$ & $0$ & $0$ & $0$ & $-C_A\frac{4 y^2(1-\kappa) \left(1+\kappa^2\right)  }{\left(1-\kappa^2 y^2\right)^2}$ \\
       $q,L$     & $0$ & $0$ & $C_F\frac{2y (1-\kappa)  }{1-\kappa^2 y^2}$ & $n_f\frac{4 y (1-\kappa) }{\left(1-\kappa^2 y^2\right)^2}$ & $0$ & $n_f\frac{4y (1-\kappa)   }{\left(1-\kappa^2 y^2\right)^2}$ \\
       $g,L$     & $0$ & $0$ & $-C_F\frac{4 y (1-\kappa)  }{1-\kappa^2 y^2}$ & $-C_A\frac{8 y (1-\kappa)  }{\left(1-\kappa^2 y^2\right)^2}$ & $0$ & $-C_A\frac{8 y (1-\kappa)  }{\left(1-\kappa^2 y^2\right)^2}$ \\
       $q,T$     & $0$ & $0$ & $0$ & $0$ & $0$ & $0$ \\
       $g,T$     & $-C_F\frac{4 (1-\kappa) }{1-\kappa^2 y^2}$ & $-C_A\frac{4 (1-\kappa) \left(1+\kappa^2 y^2\right) }{\left(1-\kappa^2 y^2\right)^2}$ & $C_F\frac{4 (1-\kappa) }{1-\kappa^2 y^2}$ & $C_A\frac{8 (1-\kappa)}{\left(1-\kappa^2 y^2\right)^2}$ & $0$ & $-C_A\frac{8 \kappa^2 y^2 (1-\kappa)}{\left(1-\kappa^2 y^2\right)^2}$ \\ 
    \end{tabular}}
    
    \caption{Same as Tab.~\ref{tab_R1_definitive} but for
    $R^{Y/\Gamma,+,[1]}_{2,i/j}$.}
    \label{tab_R2_definitive}
\end{table}

\begin{table}[!ht]
  \centering\resizebox{\textwidth}{!}{
    \begin{tabular}{c||c|c|c|c|c|c}
      \diagbox{$i,Y$}{$j,\Gamma$}& $q,U$ & $g,U$ & $q,L$ & $g,L$ & $q,T$ & $g,T$ \\
     \hhline{=#=|=|=|=|=|=}
       $q,U$      & $C_F\frac{2}{\kappa(1+\kappa y)}$ & $-n_f\frac{ 1-\kappa  y }{\kappa  (1+\kappa  y)^2}$ & $0$ & $0$ & $0$ & $-n_f\frac{1-\kappa  y}{\kappa  (1+\kappa  y)^2}$ \\
       $g,U$      & $-C_F\frac{2   \left(1-\kappa ^2\right)}{\kappa  (1+\kappa  y)}$ & $ C_A\frac{1}{1+\kappa y} \qa\frac{4\kappa}{1+\kappa y} - \frac{1+\kappa^2}{\kappa}\qc$ & $0$ & $0$ & $0$ & $C_A\frac{  \left(1+\kappa ^2\right) (1-\kappa  y)}{\kappa  (1+\kappa  y)^2}$ \\
       $q,L$      & $0$ & $0$ & $\frac{2  C_F}{1+\kappa  y}$ & $n_f\frac{3+\kappa  y}{(1+\kappa  y)^2}$ & $0$ & $-n_f\frac{ 1-\kappa  y }{\kappa  (1+\kappa  y)^2}$ \\
       $g,L$      & $0$ & $0$ & $-\frac{4  C_F}{1+\kappa  y}$ & $-C_A\frac{2  (3+\kappa  y)}{(1+\kappa  y)^2}$ & $0$ & $C_A\frac{2  (1-\kappa  y) }{\kappa  (1+\kappa  y)^2}$ \\
       $q,T$      & $0$ & $0$ & $0$ & $0$ & $0$ & $0$ \\
       $g,T$      & $-C_F\frac{4  }{y(1+\kappa  y)}$ & $C_A\qa \frac{2\kappa}{1+\kappa y} + \frac{4\kappa}{(1+\kappa y)^2} - \frac{4}{y}\qc$ & $\frac{4  C_F}{1+\kappa  y}$ & $C_A\frac{2   (3+\kappa  y)}{(1+\kappa  y)^2}$ & $0$ & $C_A\frac{2   \kappa  (1-\kappa  y)}{(1+\kappa  y)^2}$ \\
    \end{tabular}}

    \caption{Same as Tab.~\ref{tab_R1_definitive} but for
    $\widehat{R}^{Y/\Gamma,+,[1]}_{i/j}$.}
    \label{tab_Rhat_definitive}
\end{table}

\begin{table}[!ht]
  \centering
    \begin{tabular}{c||c|c|c|c|c|c}
      \diagbox{$i,Y$}{$j,\Gamma$}& $q,U$ & $g,U$ & $q,L$ & $g,L$ & $q,T$ & $g,T$ \\
     \hhline{=#=|=|=|=|=|=}
       $q,U$     & $0$ & $\frac{n_f}{\kappa}$ & $0$ & $0$ & $0$ & $\frac{n_f}{\kappa}$ \\
       $g,U$     & $0$ & $-\frac{\left(1+\kappa^2\right) C_A}{\kappa}$ & $0$ & $0$ & $0$ & $-\frac{\left(1+\kappa^2\right) C_A}{\kappa}$ \\
       $q,L$     & $0$ & $0$ & $0$ & $n_f$ & $0$ & $\frac{n_f}{\kappa}$ \\
       $g,L$     & $0$ & $0$ & $0$ & $-2C_A$ & $0$ & $-\frac{2C_A}{\kappa}$ \\
       $q,T$     & $0$ & $0$ & $0$ & $0$ & $0$ & $0$ \\
       $g,T$     & $0$ & $-2\kappa C_A$ & $0$ & $2C_A$ & $0$ & $-2\kappa C_A$ \\ 
    \end{tabular}

    \caption{Same as Tab.~\ref{tab_R1_definitive} but for
    $\Sigma^{Y/\Gamma,+,[1]}_{i/j}$.}
    \label{tab_Sigma}
\end{table}

We conclude this section by discussing the forward limit of the matching
functions and corresponding parton distributions, which amounts to taking the limits $\kappa\rightarrow0, \Delta_T\to 0$. Given the
decompositions in Eqs.~(\ref{eq:MatchingDecompositionforNonSinglet})
and~(\ref{eq:MatchingDecompositionforSinglet}), we only need to consider the
limit of the functions $R^{Y/\Gamma,-,[1]}_{1}$ and
$R^{Y/\Gamma,+,[1]}_{1,i/j}$. Moreover, since
$R^{Y/\Gamma,-,[1]}_{1}(y,0)=R^{Y/\Gamma,+,[1]}_{1,q/q}(y,0)$, it is enough to
consider the forward limit of the latter set of functions, which is collected in
Tab.~\ref{tab_R1_definitive_forward}. We first notice that the known results for
the TMD matching functions are consistently recovered (see, \textit{e.g.},
Ref.~\cite{Echevarria:2016scs, Gutierrez-Reyes:2017glx}). We also point out that
in Tab.~\ref{tab_R1_definitive_forward} we removed the rightmost column
corresponding to the linearly-polarised gluon PDF (\textit{i.e.} the forward
limit of the linearly-polarised gluon GPD). The reason is that this distribution
can only exist in hadrons with spin equal to or larger than
one~\cite{Jaffe:1989xy, Vogelsang:1998yd}. Since here we are only concerned with
the proton (see below), we can neglect this distribution. Therefore, even those
matching functions associated with the linearly-polarised gluon PDFs which
survive the forward limit, such as $R^{U/T,+,[1]}_{1,q/g}$ and
$R^{U/T,+,[1]}_{1,g/g}$, would give a vanishing contribution to the
corresponding TMD.

\begin{table}[!ht]
  \centering
    \begin{tabular}{c||c|c|c|c|c}
      \diagbox{$i,Y$}{$j,\Gamma$}& $q,U$ & $g,U$ & $q,L$ & $g,L$ & $q,T$ \\
     \hhline{=#=|=|=|=|=}
       $q,U$     & $2C_F(1-y) $ & $4n_f  y(1-y) $ & $0$ & $0$ & $0$  \\
       $g,U$     & $2 C_F y$ & $0$ & $0$ & $0$ & $0$  \\
       $q,L$     & $0$ & $0$ & $2C_F(1-y)$ & $4n_f (1-y)$ & $0$  \\
       $g,L$     & $0$ & $0$ & $-4C_F(1-y)$ & $-8C_A(1-y)$ & $0$  \\
       $q,T$     & $0$ & $0$ & $0$ & $0$ & $0$  \\
       $g,T$     & $-4C_F\frac{1-y }{y }$ & $-4C_A \frac{1-y }{y }$ & $0$ & $0$ & $0$  
    \end{tabular}
    \caption{Forward limit (i.e. $\kappa\to 0, \Delta_T\to0$) of the functions
    $R^{Y/\Gamma,+,[1]}_{1,i/j}$ given in Tab.~\ref{tab_R1_definitive}.}
    \label{tab_R1_definitive_forward}
\end{table}
\section{Matching for specific parametrisations}
\label{sec_matching_param}

In the previous section, we obtained the matching functions for GTMDs on GPDs by
considering the appropriate matrix elements (Eqs.~(\ref{QuarkCorrelator})
and~(\ref{GluonCorrelator}) for GTMDs, and Eqs.~(\ref{QuarkGPDCorrelator})
and~(\ref{GluonGPDCorrelator}) for GPDs). Restricting to the case of a proton
target ($H=p$), we now need to adopt specific parametrisations for both GTMD and
GPD correlators and establish their matching at the level of single
distributions. To this purpose, we use a modified version of the GTMD
parametrisation presented in Ref.~\cite{Lorce:2013pza}, which in turn slightly
differs from that of Ref.~\cite{Meissner:2009ww}. In the following, we quickly
review the logic behind its construction. The easiest way to derive the matching
relations between GTMDs and GPDs is to use the helicity basis for the proton
state in which the matrix elements in Eqs.~(\ref{QuarkCorrelator})
and~(\ref{GluonCorrelator}) can generally be written as: 
\begin{align}\label{MatrixElement}
\mathcal{F}^{[Y]}_{i/H}(x,\xi,\bm{b},\bm{\Delta}_T)=\bar{u}_{\lambda'}(p') \frac{M_i^{Y}(x,\xi,\bm{b},\bm{\Delta}_T)}{2p^+(1-\xi^2)} u_\lambda(p),
\end{align}
with $u$ and $\overline{u}$ being the spinors of  incoming and outgoing protons,
respectively. Dropping all arguments, the decompositions for polarisations $U$
and $L$ is the same for quarks and gluons and read:
\begin{align}\label{eq:ParemtrizationUnpolGTMD}
M_i^{U}&= \gamma^+ \ta S_{1,1a}^{0;+;i} + \gamma_5 \frac{i\e_T^{b\Delta_T}}{M|\bm{b}|}S_{1,1b}^{0;+;i}\tc -i\sigma^{\mu+} \ta \frac{b_\mu}{|\bm{b}|} P_{1,1a}^{0;+;i} + \frac{\Delta_{T,\mu}}{M} P_{1,1b}^{0;+;i}\tc\,, \\
\label{eq:ParemtrizationLongpolGTMD}M_i^{L}&= \gamma^+\gamma_5 \ta S_{1,1a}^{0;-;i} +\gamma_5 \frac{i\e_T^{b\Delta_T}}{M|\bm{b}|} S_{1,1b}^{0;-;i}\tc -i\sigma^{\mu+}\gamma_5 \ta \frac{b_\mu}{|\bm{b}|} P_{1,1a}^{0;-;i} + \frac{\Delta_{T,\mu}}{M} P_{1,1b}^{0;-;i}\tc\,,
\end{align}
with $i=q,g$ and where $M$ is the proton mass. For polarisation $T$, quark and
gluon decompositions are different and read:
\begin{equation}
\label{eq:ParemtrizationQuarkTransGTMD}
\begin{split}M_q^{T} =& \sqrt{2}\,i\e^{R\mu}_T\gamma^+ \ta b_\mu M P_{1,1a}^{1;-;q}+\frac{\Delta_{T,\mu}}{M} P_{1,1b}^{1;-;q}\tc+\sqrt{2}\gamma^+\gamma_5 \ta b_R M P_{1,1a}^{'1;-;q}+\frac{\Delta_R}{M} P_{1,1b}^{'1;-;q}\tc \\
& + \frac{\sqrt{2}\,i\sigma^{R+}\gamma_5}{2}\ta S_{1,1a}^{1;-;q} - \gamma_5 i\e_T^{b\Delta_T}S_{1,1b}^{1;-;q} \tc +\sqrt{2}\,i\sigma^{L+}\gamma_5\ta \frac{b_R^2}{\bm{b}^2} D_{1,1a}^{1;-;q}+ \frac{\Delta_R^2}{M^2}D_{1,1b}^{1;-;q}\tc\,,
\end{split}
\end{equation}
\begin{align}
\label{eq:ParemtrizationGluonsTransGTMD}M_g^{T} &=2\gamma^+ \ta \frac{b_R^2}{\bm{b}^2} D_{1,1a}^{2;+;g}+ \frac{\Delta_R^2}{M^2}D_{1,1b}^{2;+;g}\tc - 2i\e^{R\mu}_T\gamma^+\gamma_5   \ta  \frac{b_\mu b_R}{\bm{b}^2}D_{1,1a}^{'2;+;g} + \frac{\Delta_{T,\mu}\Delta_R}{M^2}D_{1,1b}^{'2;+;g} \tc \notag\\
& - i\sigma^{R+} \ta \frac{b_R}{|\bm{b}|} P_{1,1a}^{2;+;g} + \frac{\Delta_R}{M} P_{1,1b}^{2;+;g}\tc + 2i\sigma^{L+} \ta \frac{b_R^2\Delta_R}{M\bm{b}^2} F_{1,1a}^{'2;+;g} + \frac{b_R\Delta_R^2}{M^2|\bm{b}|}F_{1,1b}^{'2;+;g} \tc\,.
\end{align}
As mentioned above, this is a modified version of the parametrisation of
Ref.~\cite{Lorce:2013pza} in different respects. First,
Ref.~\cite{Lorce:2013pza} provides expressions in momentum space. This is
inconvenient for matching, which takes place naturally in position space. We
thus do the replacement $k_T^\mu/M \to b^\mu/|\bf{b}|$, but keeping the same
structures. Second, we replace the factor $\sqrt{1-\xi^2}$ in Eq.~(3.52) of
Ref.~\cite{Lorce:2013pza} with $(1-\xi^2)$, see Eq.~(\ref{MatrixElement}). The
net result is that $\mathcal{F}_{\,\text{here}} = \mathcal{F}_{\text{\tiny
\cite{Lorce:2013pza}}} / \sqrt{1-\xi^2}$. This allows us to get rid of an
overall factor of $\sqrt{1-\xi^2}$ in the matching functions. Finally, the
distributions $F_{1,1a}^{'2;+;g}$ and $F_{1,1b}^{'2;+;g}$ appearing in the
decomposition of $M_g^{T}$ could in principle be written as the following
combinations of the distributions $F_{1,1a}^{2;+;g}$ and $F_{1,1b}^{2;+;g}$ of
Ref.~\cite{Lorce:2013pza}:
\begin{align}
    F_{1,1a}^{'2;+;g} &=\frac{1}{\sqrt{1-\xi^2}}\left[\frac{2 M^2 (\bm{b}\cdot\bm{\Delta}_T)}{\bm{\Delta}_T^2}F_{1,1a}^{2;+;g}-\frac{\bm{\Delta}_T^2}{M^2}F_{1,1b}^{2;+;g}\right]\,,\\
    F_{1,1b}^{'2;+;g} &=\frac{1}{\sqrt{1-\xi^2}}\left[\frac{2 (\bm{b}\cdot\bm{\Delta}_T)}{M |\bm{b}|}F_{1,1b}^{2;+;g}-\frac{M^3|\bm{b}|}{\bm{\Delta}_T^2}F_{1,1a}^{2;+;g}\right]\,.
\end{align} 
However, the tensor structures in front of the non-primed distributions do not
naturally emerge from the matching, leading to needlessly complex expressions.
In what follows, we will thus only consider $F_{1,1a}^{'2;+;g}$ and
$F_{1,1b}^{'2;+;g}$, rather than their non-primed counterparts.

As far as GPDs are concerned, we use the standard parametrisation of
Ref.~\cite{Meissner:2009ww}, where the leading-twist quark and gluon
distributions
$H^i,E^i,\widetilde{H}^i,\widetilde{E}^i,H_T^i,E^i_T,\widetilde{H}_T^i,\widetilde{E}_T^i$,
with $i=q,g$, are defined.

We are now in a position to give the matching of GTMDs on GPDs for each
individual distribution. Unpolarised quark GTMD distributions read:
\begin{align}
\label{S^0+q_11a}S_{1,1a}^{0;+;q} &= \mathcal{C}_{q/q}^{U/U}\otimes\qa(1-\xi^2) H^q -\xi^2E^q\qc+\mathcal{C}_{q/g}^{U/U}\otimes \qa (1-\xi^2)H^g -\xi^2E^g\qc\\
&\notag   - \frac{2(\bm{b}\cdot\bm{\Delta}_T)^2-\bm{b}^2\bm{\Delta}_T^2}{4M^2\bm{b}^2}\mathcal{C}_{q/g}^{U/T}\otimes\qa E_T^g -\xi \widetilde{E}^g_T+2\widetilde{H}^g_T\qc\,,\\
\label{eq_S11b0pq_res}
S_{1,1b}^{0;+;q} &=\frac{(\bm{b}\cdot\bm{\Delta}_T)}{2M|\bm{b}|} \mathcal{C}_{q/g}^{U/T}\otimes\qa \xi E_T^g - \widetilde{E}^g_T\qc\,,\\
P_{1,1a}^{0;+;q} &= \frac{(\bm{b}\cdot\bm{\Delta}_T)}{M|\bm{b}|} \mathcal{C}_{q/g}^{U/T}\otimes\qa(1-\xi^2)H_T^g+\xi \widetilde{E}_T^g-\xi^2E_T^g\qc\,,\\
P_{1,1b}^{0;+;q} &=\frac{1}{2} \ta \mathcal{C}_{q/q}^{U/U}\otimes E^q + \mathcal{C}_{q/g}^{U/U}\otimes E^g\tc \\
& \notag -\frac{1}{2}\mathcal{C}_{q/g}^{U/T} \otimes\qa (1-\xi^2)H_T^g+\xi \widetilde{E}_T^g-\xi^2E_T^g-\frac{2(\bm{b}\cdot\bm{\Delta}_T)^2-\bm{b}^2\bm{\Delta}_T^2}{2M^2\bm{b}^2}\widetilde{H}_T^g\qc\,.
\end{align}
Longitudinally-polarised quark GTMDs read:
\begin{align}
S_{1,1a}^{0;-;q} &= \mathcal{C}_{q/q}^{L/L}\otimes \qa (1-\xi^2)\widetilde{H}^q -\xi^2\widetilde{E}^q\qc + \mathcal{C}_{q/g}^{L/L}\otimes \qa (1-\xi^2)\widetilde{H}^g -\xi^2\widetilde{E}^g\qc \\
& \notag +\frac{2(\bm{b}\cdot\bm{\Delta}_T)^2-\bm{b}^2\bm{\Delta}_T^2}{4M^2\bm{b}^2} \mathcal{C}_{q/g}^{L/T}\otimes \qa \widetilde{E}_T^g-\xi E_T^g\qc\,,\\
S_{1,1b}^{0;-;q} &=\frac{(\bm{b}\cdot\bm{\Delta}_T)}{2M|\bm{b}|} \mathcal{C}_{q/g}^{L/T}\otimes\qa E_T^g -\xi \widetilde{E}_T^g+2\widetilde{H}_T^g\qc\,,\\
P_{1,1a}^{0;-;q} &=\frac{(\bm{b}\cdot\bm{\Delta}_T)}{M|\bm{b}|} \mathcal{C}_{q/g}^{L/T}\otimes\qa (1-\xi^2)H_T^g+\xi \widetilde{E}_T^g-\xi^2E_T^g + \frac{\bm{\Delta}_T^2}{2M^2}\widetilde{H}_T^g\qc\,, \\
\label{P^0-q_11b}P_{1,1b}^{0;-;q} &= \frac{1}{2}\ta \mathcal{C}_{q/q}^{L/L}\otimes [\xi\widetilde{E}^q] + \mathcal{C}_{q/g}^{L/L}\otimes [\xi\widetilde{E}^g] \tc \\
&\notag - \frac{1}{2}\mathcal{C}_{q/g}^{L/T}\otimes\qa (1-\xi^2) H_T^g+\xi \widetilde{E}_T^g-\xi^2E_T^g+\frac{(\bm{b}\cdot\bm{\Delta}_T)^2}{M^2\bm{b}^2}\widetilde{H}_T^g\qc\,.
\end{align}
Transversely-polarised quark GTMDs read:
\begin{align}
\label{S^1-q_11a}S_{1,1a}^{1;-;q} &=2 \mathcal{C}_{q/q}^{T/T}\otimes\qa  (1-\xi^2) H_T^q + \xi\widetilde{E}_T^q-\xi^2 E_T^q + \frac{\bm{\Delta}_T^2}{4M^2}\widetilde{H}_T^q \qc\,,  \\
P_{1,1b}^{1;-;q} &=\frac{1}{2} \mathcal{C}_{q/q}^{T/T}\otimes\qa E_T^q-\xi \widetilde{E}_T^q+2\widetilde{H}_T^q\qc\,,\\
P_{1,1b}^{'1;-;q} &= \frac{1}{2}\mathcal{C}_{q/q}^{T/T}\otimes\qa\widetilde{E}_T^q-\xi E_T^q\qc\,,\\
\label{D^1-q_11b}D_{1,1b}^{1;-;q} &= -\frac{1}{2}\mathcal{C}_{q/q}^{T/T}\otimes\widetilde{H}_T^q\,.
\end{align}

For unpolarised and longitudinally-polarised gluon GTMDs, the structure of the
matching is analogous to that of quarks. Specifically, for unpolarised gluons
GTMDs we have:
\begin{align}
\label{S^0+g_11a}S_{1,1a}^{0;+;g} &= \mathcal{C}_{g/g}^{U/U}\otimes\qa(1-\xi^2) H^g -\xi^2E^g\qc+\mathcal{C}_{g/q}^{U/U}\otimes \qa (1-\xi^2)H^q -\xi^2E^q\qc \\
&\notag- \frac{2(\bm{b}\cdot\bm{\Delta}_T)^2-\bm{b}^2\bm{\Delta}_T^2}{4M^2\bm{b}^2}\mathcal{C}_{g/g}^{U/T}\otimes\qa E_T^g -\xi \widetilde{E}^g_T+2\widetilde{H}^g_T\qc\,,\\
\label{eq_S11b0pg_res}
S_{1,1b}^{0;+;g} &=\frac{(\bm{b}\cdot\bm{\Delta}_T)}{2M|\bm{b}|} \mathcal{C}_{g/g}^{U/T}\otimes\qa \xi E_T^g - \widetilde{E}^g_T\qc\,,\\
P_{1,1a}^{0;+;g} &= \frac{(\bm{b}\cdot\bm{\Delta}_T)}{M|\bm{b}|} \mathcal{C}_{g/g}^{U/T}\otimes\qa(1-\xi^2)H_T^g+\xi \widetilde{E}_T^g-\xi^2E_T^g\qc\,,\\
P_{1,1b}^{0;+;g} &=\frac{1}{2} \ta \mathcal{C}_{g/g}^{U/U}\otimes E^g + \mathcal{C}_{g/q}^{U/U}\otimes E^q \tc \\
& \notag -\frac{1}{2}\mathcal{C}_{g/g}^{U/T} \otimes\qa (1-\xi^2)H_T^g+\xi \widetilde{E}_T^g-\xi^2E_T^g-\frac{2(\bm{b}\cdot\bm{\Delta}_T)^2-\bm{b}^2\bm{\Delta}_T^2}{2M^2\bm{b}^2}\widetilde{H}_T^g\qc\,,
\end{align}
and for longitudinally-polarised gluon GTMDs, we have: 
\begin{align}
S_{1,1a}^{0;-;g} &= \mathcal{C}_{g/g}^{L/L}\otimes \qa (1-\xi^2)\widetilde{H}^g -\xi^2\widetilde{E}^g\qc + \mathcal{C}_{g/q}^{L/L}\otimes \qa (1-\xi^2)\widetilde{H}^q -\xi^2\widetilde{E}^q\qc \\
& \notag +\frac{2(\bm{b}\cdot\bm{\Delta}_T)^2-\bm{b}^2\bm{\Delta}_T^2}{4M^2\bm{b}^2} \mathcal{C}_{g/g}^{L/T}\otimes \qa \widetilde{E}_T^g-\xi E_T^g\qc\,,\\
S_{1,1b}^{0;-;g} &=\frac{(\bm{b}\cdot\bm{\Delta}_T)}{2M|\bm{b}|} \mathcal{C}_{g/g}^{L/T}\otimes\qa E_T^g -\xi \widetilde{E}_T^g+2\widetilde{H}_T^g\qc\,,\\
P_{1,1a}^{0;-;g} &=\frac{(\bm{b}\cdot\bm{\Delta}_T)}{M|\bm{b}|} \mathcal{C}_{g/g}^{L/T}\otimes\qa (1-\xi^2)H_T^g+\xi \widetilde{E}_T^g-\xi^2E_T^g + \frac{\bm{\Delta}_T^2}{2M^2}\widetilde{H}_T^g\qc\,,\\
P_{1,1b}^{0;-;g} &=\frac{1}{2}\ta \mathcal{C}_{g/g}^{L/L}\otimes [\xi\widetilde{E^g}] + \mathcal{C}_{g/q}^{L/L}\otimes [\xi\widetilde{E}^q]\tc \\
& \notag - \frac{1}{2}\mathcal{C}_{g/g}^{L/T}\otimes\qa (1-\xi^2) H_T^g+\xi \widetilde{E}_T^g-\xi^2E_T^g+\frac{(\bm{b}\cdot\bm{\Delta}_T)^2}{M^2\bm{b}^2}\widetilde{H}_T^g\qc\,.
\end{align}

For the linearly-polarised gluons GTMDs, the parametrisation differs from that
of quarks and the picture is more complicated. In particular, there are eight
distributions that match onto GPDs and they are given by:
\begin{align}
P_{1,1a}^{2;+;g} &=-\frac{(\bm{b}\cdot\bm{\Delta}_T)}{M|\bm{b}|} \ta \mathcal{C}_{g/g}^{T/U}\otimes E^g - \mathcal{C}_{g/g}^{T/L} \otimes [\xi\widetilde{E}^g] + \mathcal{C}_{g/q}^{T/U} \otimes E^q - \mathcal{C}_{g/q}^{T/L}\otimes [\xi\widetilde{E}^q]\tc\,,\\
P_{1,1b}^{2;+;g} &= -\mathcal{C}_{g/g}^{T/T} \otimes\qa (1-\xi^2)H_T^g+\xi \widetilde{E}_T^g-\xi^2 E_T^g+\frac{\bm{\Delta}_T^2}{4M^2}\widetilde{H}^g_T\qc \\ 
\notag & +\frac{1}{2} \mathcal{C}_{g/g}^{T/U} \otimes E^g-\frac{1}{2}\mathcal{C}_{g/g}^{T/L}\otimes[\xi\widetilde{E}^g]
+\frac{1}{2}\mathcal{C}_{g/q}^{T/U}\otimes E^q-\frac{1}{2}\mathcal{C}_{g/q}^{T/L}\otimes[\xi\widetilde{E}^q]\,,\\
D_{1,1a}^{2;+;g} &= \mathcal{C}_{g/q}^{T/U}\otimes \qa(1-\xi^2) H^q -\xi^2 E^q\qc + \mathcal{C}_{g/g}^{T/U} \otimes\qa(1-\xi^2) H^g -\xi^2 E^g\qc\,, \\
D_{1,1b}^{2;+;g} &=-\frac{1}{4} \mathcal{C}_{g/g}^{T/T}\otimes \qa E^g_T-\xi\widetilde{E}^g_T+2\widetilde{H}^g_T\qc\,, \\
D_{1,1a}^{'2;+;g} &= \mathcal{C}_{g/q}^{T/L} \otimes\qa (1-\xi^2)\widetilde{H}^q -\xi^2 \widetilde{E}^q\qc + \mathcal{C}_{g/g}^{T/L} \otimes\qa (1-\xi^2)\widetilde{H}^g -\xi^2 \widetilde{E}^g\qc\,,\\
D_{1,1b}^{'2;+;g} &=-\frac{1}{4}\mathcal{C}_{g/g}^{T/T}\otimes \qa\widetilde{E}^g_T-\xi E^g_T\qc\,, \\
F_{1,1a}^{'2;+;g} &=-\frac{\bm{\Delta}_T^2}{4M^2}  \mathcal{C}_{g/g}^{T/T} \otimes\widetilde{H}_T^g+\frac{1}{2}\mathcal{C}_{g/g}^{T/U}\otimes E^g+\frac{1}{2}\mathcal{C}_{g/g}^{T/L}\otimes [\xi\widetilde{E}^g] \\
\notag &+ \frac{1}{2}\mathcal{C}_{g/q}^{T/U}\otimes E^q+\frac{1}{2}\mathcal{C}_{g/q}^{T/L}\otimes[\xi\widetilde{E}^q]\,,\\
\label{F^'2+g_11b}F_{1,1b}^{'2;+;g} &= \frac{(\bm{b}\cdot\bm{\Delta}_T)}{2M|\bm{b}|}\mathcal{C}_{g/g}^{T/T}\otimes\widetilde{H}_T^g\,.
\end{align}

We remark that, thanks to the dependence on the scalar product
$\bm{b}\cdot\bm{\Delta}_T$, we find several non-vanishing GTMDs upon matching on
twist-two GPDs. This is a consequence of the non-forward nature of GTMDs, which
provides a second transverse vector that can couple to $\bm{b}$ to generate
angular modulations already at leading twist.

\section{Reconstructing GTMDs}
\label{sec_reconstructingGTMDs}

Having established the matching pattern of leading-twist GTMDs on GPDs, we turn
to their evolution. This will complete the picture and allow us to reconstruct
the full kinematics dependence of GTMDs. To discuss evolution, it is convenient
to denote with $\mathbb{F}_{i}^{[Y]}$ any quark ($i=q$) or gluon ($i=g$) GTMD
from an explicit parametrisation of polarisation $Y$. The evolution equations
with respect to the rapidity scale $\zeta$ and the renormalisation scale $\mu$
read:
\begin{equation}
  \begin{array}{rcl}
    \displaystyle
    \frac{d\log\mathbb{F}_{i}^{[Y]} (x,\xi,\bm{b},\bm{\Delta}_T;\mu,\zeta)}{d\log\sqrt{\zeta}}&=&\displaystyle
    K_i(\bm{b},\mu_b) - \int_{\mu_b}^{\mu}
    \frac{d\mu'}{\mu'} \gamma_{K,i}(\alpha_s(\mu'))\,,\\
    \\
    \displaystyle\frac{d\log\mathbb{F}_{i}^{[Y]} (x,\xi,\bm{b},\bm{\Delta}_T;\mu,\zeta)}{d\log\mu}&=&\displaystyle \gamma_{F,i}(\alpha_s(\mu)) - \frac{\gamma_{K,i}(\alpha_s(\mu))}{2}\ta \log\left(\frac{|1-\kappa^2|\zeta}{\mu}\right)  + is\pi \theta(\kappa-1)\tc\,,
  \end{array}
  \label{eq:evolutionequationsExp}
\end{equation}
with $\mu_{b}=b_0/|\bm{b}|$ and where the anomalous dimensions
$K_i(\bm{b},\mu_b)$, $\gamma_{F,i}$, and $\gamma_{K,i}$ are all perturbative
quantities and coincide with their forward TMD counterparts. In fact, the main
differences between GTMD and TMD are the presence in the evolution equation
w.r.t. $\mu$ of the factor $|1-\kappa^2|$ inside the logarithm that multiplies
$\gamma_{K,i}$ and of the imaginary part, which, as we will see below, plays a
crucial role.\footnote{As an aside note, we point out that a similar factor
emerges in the evolution of twist-3 forward TMDs, see
Ref.~\cite{Rodini:2022wki}. This happens because in both cases one has two
independent longitudinal-momentum variables: for GTMDs, they are $x$ and $\xi$,
while for twist-3 TMDs they are any pair of longitudinal-momentum fractions of
the three fields that define the operator. In both cases, the imaginary part in
the evolution equations comes with the sign of the direction of the Wilson line
$s$.}

The evolution equations in Eq.~(\ref{eq:evolutionequationsExp}) can be solved
explicitly. Choosing $\mu_0=\mu_b$ and $\zeta_0=\mu_b^2$ as boundary-condition
scales, which prevent the presence of potentially large logarithms, the
evolution of GTMDs to the scales $\mu$ and $\zeta$ is multiplicative and can be
written as:
\begin{equation}
  \mathbb{F}_{i}^{[Y]} (x,\xi,\bm{b},\bm{\Delta}_T;\mu,\zeta)=\mathcal{R}_i\left[(\mu,\zeta)\leftarrow
    (\mu_{b},\mu_{b}^2)\right]\mathbb{F}_{i}^{[Y]}
  (x,\xi,\bm{b},\bm{\Delta}_T;\mu_b,\mu_b^2)\,,
\label{eq:GTMDevolutionSolved}
\end{equation}
where the factor $\mathcal{R}_i$, often referred to as Sudakov form factor, has
the following form:
\begin{equation}
\begin{array}{rcl}
  \displaystyle  \mathcal{R}_i\left[(\mu,\zeta)\leftarrow
  (\mu_{b},\mu_{b}^2)\right] &=&\displaystyle \exp
                                     \bigg\{\frac{1}{2} K_i(\bm{b},\mu_{b}) \log \left(\frac{|1-\kappa^2|\zeta}{\mu_{b}^2}\right) 
                                    \\
  \\
                                 &+& \displaystyle
                                     \int_{\mu_{b}}^{\mu} \frac{d\mu'}{\mu'}\left[ \gamma_{F,i}
                                     (\alpha_s(\mu')) - \frac{1}{2}\gamma_{K,i} (\alpha_s(\mu')) \log
                                     \left(\frac{|1-\kappa^2|\zeta}{\mu'^2}\right) \right] \\
    \\
                                &+& \displaystyle \frac{i\pi s }{2} \theta(\kappa-1) \qa K_i(\bm{b},\mu_{b}) - \int_{\mu_{b}}^{\mu} \frac{d\mu'}{\mu'}\gamma_{K,i} (\alpha_s(\mu')) \qc \bigg\} \, .
\end{array}
\label{eq:SudakovFormFactorGTMDs}
\end{equation}
Factoring out the imaginary part of the argument of the exponential in
Eq.~(\ref{eq:SudakovFormFactorGTMDs}) allows us to write:
\begin{align}
\label{eq:SudakovFormFactorGTMDsDecomp}
& \displaystyle  \mathcal{R}_i\left[(\mu,\zeta)\leftarrow
  (\mu_{b},\mu_{b}^2)\right] = \\
&\notag \quad =R_i\left[(\mu,\zeta)\leftarrow
  (\mu_{b},\mu_{b}^2)\right]\exp
   \bigg\{s\frac{i\pi}{2}\theta(\kappa-1) \left[
  K_i(\bm{b},\mu_{b}) -
 \int_{\mu_{b}}^{\mu}
 \frac{d\mu'}{\mu'} \gamma_{K,i}
                                 (\alpha_s(\mu')) \right] \bigg\} \\
& \notag \quad = R_i\left[(\mu,\zeta)\leftarrow
  (\mu_{b},\mu_{b}^2)\right]\left[\cos\left(\phi(\kappa,\bm{b},\mu)\right)+i s \sin\left(\phi(\kappa,\bm{b},\mu)\right)\right]\,,
\end{align}
where $R_i$ is real and we have defined the angle:
\[
\phi(\kappa,\bm{b},\mu) = \frac{\pi}{2}\theta(\kappa-1)\left(K_i(\bm{b},\mu_{b}) -
\int_{\mu_{b}}^{\mu}
\frac{d\mu'}{\mu'} \gamma_{K,i}(\alpha_s(\mu'))\right)=\frac{\pi}{2}\theta(\kappa-1)
    K_i(\bm{b},\mu)\,,
\]
where $K_i(\bm{b},\mu)$, as opposed to $K_i(\bm{b},\mu_{b})$, is the
\textit{evolved} Collin-Soper kernel which governs the rapidity-scale evolution
in Eq.~(\ref{eq:evolutionequationsExp}).

Now, we use the fact that each GTMD can be decomposed into a T-even and a T-odd
part as follows:
\begin{equation}
  \mathbb{F}_{i}^{[Y]}=\mathbb{F}_{i}^{[Y],e} + i\mathbb{F}_{i}^{[Y],o}\,,
\end{equation}
where $\mathbb{F}_{i}^{[Y],e}$ and $\mathbb{F}_{i}^{[Y],o}$ are both real
functions. Moreover, $\mathbb{F}_{i}^{[Y],e}$ remains unchanged upon change of
the direction of the Wilson line $s$, while $\mathbb{F}_{i}^{[Y],o}$ changes
sign. Dropping all variables except the scales,
Eq.~(\ref{eq:GTMDevolutionSolved}) can be expressed in a matrix form for T-even
and T-odd components as follows:
\begin{align}
\label{eq:BeforeMatching}
&\begin{pmatrix}
\mathbb{F}_{i}^{[Y],e} (\mu,\zeta)\\
\mathbb{F}_{i}^{[Y],o} (\mu,\zeta)
\end{pmatrix}   = \displaystyle
R_i\left[(\mu,\zeta)\leftarrow
  (\mu_{b},\mu_{b}^2)\right]
\begin{pmatrix}
\cos\left(\phi(\mu) \right) & -s\sin\left(\phi(\mu) \right)\\
s\sin\left(\phi(\mu) \right) & \cos\left(\phi(\mu) \right)
\end{pmatrix}\displaystyle
\begin{pmatrix}
\mathbb{F}_{i}^{[Y],e} (\mu_b, \mu_b^2)\\
\mathbb{F}_{i}^{[Y],o} (\mu_b,\mu_b^2)
\end{pmatrix}\,.
\end{align}We note that the rotation
matrix generated by the Sudakov form factor is diagonal in the DGLAP region
($\kappa<1$) such that, as expected, T-even and T-odd GTMDs evolve independently
in this region. Conversely, in the ERBL region ($\kappa>1$), the rotation matrix
does cause a mixing of the two distributions upon evolution.

The ``initial-scale'' GTMD $\mathbb{F}_{i}^{[Y]}(\mu_b, \mu_b^2)$ can now be
obtained by means of matching on a suitable combination of collinear GPDs
$f_j^{[\Gamma]}(\mu_b)$ through convolution with the appropriate matching
functions, see Eq.~(\ref{MatchingOPE}):\footnote{Note that setting $\mu=\mu_b$
in Eq.~(\ref{MatchingOPE}) causes all explicit logarithms in the one-loop
matching functions in Eq.~(\ref{DecompositionMatchingCoefficient}) to vanish.}
\begin{equation}
  \mathbb{F}_{i}^{[Y]}(\mu_b,\mu_b^2)=\left[\mathcal{C}^{Y/\Gamma}_{i/j}\otimes f_j^{[\Gamma]}\right](\mu_b,\mu_b^2)\,.
  \label{eq:InitialScaleMatching}
\end{equation}
Given a particular GTMD $\mathbb{F}_{i}^{[Y]}$, the appropriate set of GPDs
$f_j^{[\Gamma]}$ that enter the r.h.s. of Eq.~(\ref{eq:InitialScaleMatching})
can be read from Eqs.~(\ref{S^0+q_11a})-(\ref{F^'2+g_11b}). As we have seen
above, the matching functions $\mathcal{C}^{Y/\Gamma}_{i/j}$ also acquire an
imaginary contribution (see Eq.~(\ref{eq:MatchingDecompositionforSinglet})), so
that one can write:\footnote{Note that an imaginary part is only present in the
functions $\mathcal{C}_{i/g}^{Y/\Gamma}$, while $\mathcal{C}_{i/q}^{Y/\Gamma}$
are real (see Tab.~\ref{tab_Sigma}). In other words, the imaginary part of the
GTMDs generated by matching comes exclusively from gluon GPD channels.}
\begin{equation}
  \mathbb{F}^{[Y]}_{i}(\mu_b,\mu_b^2) =\left[\mathcal{C}_{i/j}^{Y/\Gamma,e}\otimes
  f^{[\Gamma]}_j\right](\mu_b,\mu_b^2)+i s\left[\mathcal{C}_{i/j}^{Y/\Gamma,o}\otimes f^{[\Gamma]}_j\right](\mu_b,\mu_b^2)\,,
\end{equation}
which in matrix form reads:
\begin{equation}
\begin{pmatrix}
\mathbb{F}_{i}^{[Y],e}(\mu_b,\mu_b^2)\\
\mathbb{F}_{i}^{[Y],o}(\mu_b,\mu_b^2)
\end{pmatrix} = \left[\begin{pmatrix}
\mathcal{C}_{i/j}^{Y/\Gamma,e}\\
s\mathcal{C}_{i/j}^{Y/\Gamma,o}
\end{pmatrix} \otimes f_j^{[\Gamma]}\right](\mu_b,\mu_b^2)\,.
\label{eq:InitialScaleMatchingMatrix}
\end{equation}
This equality can finally be used into Eq.~(\ref{eq:BeforeMatching}),
producing the result:
\begin{align}
  \label{eq:AfterMatching}
  &\begin{pmatrix}
  \mathbb{F}_{i}^{[Y],e} (\mu,\zeta)\\
  \mathbb{F}_{i}^{[Y],o} (\mu,\zeta)
  \end{pmatrix}   = \displaystyle
  R_i\left[(\mu,\zeta)\leftarrow
    (\mu_{b},\mu_{b}^2)\right]
  \begin{pmatrix}
  \cos\left(\phi(\mu) \right) & -\sin\left(\phi(\mu) \right)\\
  s\sin\left(\phi(\mu) \right) & s\cos\left(\phi(\mu) \right)
  \end{pmatrix}\displaystyle
  \left[\begin{pmatrix}
    \mathcal{C}_{i/j}^{Y/\Gamma,e}\\
    \mathcal{C}_{i/j}^{Y/\Gamma,o}
    \end{pmatrix} \otimes f_j^{[\Gamma]}\right](\mu_b,\mu_b^2)\,,
\end{align}
in which the T-parity of $\mathbb{F}_{i}^{[Y],e}$ and $\mathbb{F}_{i}^{[Y],o}$
is explicitly preserved under evolution and matching.

Finally, we point out that GTMDs in momentum (${\bm k}_T$) space can be obtained
by Fourier transforming Eq.~(\ref{eq:AfterMatching}) w.r.t. the transverse
displacement ${\bm b}$. Care must be taken when taking the Fourier transform to
include also possible $\bm b$-dependent factors that appear in the decomposition
of the relevant correlators. Therefore, the inversion of GTMDs from ${\bm b}$ to
${\bm k}_T$ space will generally read:
\begin{equation}
  \hat{\mathbb{F}}_{i}^{[Y]} (x,\xi,\bm{k}_T,\bm{\Delta}_T;\mu,\zeta) = \int d^2\bm{b}\,e^{-i\bm{b}\cdot\bm{k}_T} f({\bm b})\mathbb{F}_{i}^{[Y]} (x,\xi,\bm{b},\bm{\Delta}_T;\mu,\zeta)\,,
  \label{eq:FourierTransformGTMDGen}
\end{equation}
where the factor $f({\bm b})$ can be read off from
Eqs.~(\ref{eq:ParemtrizationUnpolGTMD}), (\ref{eq:ParemtrizationLongpolGTMD}),
(\ref{eq:ParemtrizationQuarkTransGTMD}), and
(\ref{eq:ParemtrizationGluonsTransGTMD}). More details on the numerical
implementation of the Fourier transform in
Eq.~(\ref{eq:FourierTransformGTMDGen}) for the specific GTMDs to be treated in
Sec.~\ref{sec_numerical_results} can be found in
App.~\ref{sec_Fourier_transform}.
\section{Numerical results}
\label{sec_numerical_results}

In this section, we present a numerical implementation of the results obtained
in the previous sections. In order to assess the magnitude of the T-even/T-odd
mixing effect caused by evolution (Eq.~(\ref{eq:BeforeMatching})) and matching
(Eq.~(\ref{eq:InitialScaleMatchingMatrix})), we present a selection of numerical
results for the GTMDs ${S}_{1,1a}^{0;+;q}$ and ${S}_{1,1b}^{0;+;q}$. These
distributions are defined through Eq.~(\ref{eq:ParemtrizationUnpolGTMD}) and
match on GPDs as in Eqs.~(\ref{S^0+q_11a}) and~(\ref{eq_S11b0pq_res}). Since we
will only present results in momentum space, all GTMDs are Fourier transformed
using Eq.~(\ref{eq:FourierTransformGTMDGen}) (see also
App.~\ref{sec_Fourier_transform}). Before moving to show our results, several
general remarks are in order.

As no phenomenologically accurate extractions of the non-perturbative part of
GTMDs are currently available, we resorted to using the forward TMD extraction
of Ref.~\cite{Bacchetta:2019sam}. Moreover, no reliable models for the linearly
polarised gluon GPDs $\widetilde{H}^g_T$, $E_T^g$, and $\widetilde{E}^g_T$
currently exist. In their place, we used a scaled-down version of the
corresponding unpolarised and longitudinally polarised GPDs, \textit{i.e.}
$\widetilde{H}_T^g = c \widetilde{H}^g$, $E^g_T = c E^g$, $\widetilde{E}^g_T =
c\widetilde{E}^g$ with $c$ arbitrarily chosen to be $10^{-1}$, for which we used
the Goloskokov-Kroll (GK)
model~\cite{Goloskokov:2005sd,Goloskokov:2007nt,Goloskokov:2009ia}. As far as
the perturbative accuracy is concerned, combining in
Eq.~(\ref{eq:AfterMatching}) one-loop matching functions with anomalous
dimensions computed at the appropriate perturbative order, allowed us to achieve
next-to-next-to-leading-logarithmic (NNLL) accurate
GTMDs~\cite{Bacchetta:2019sam, Bertone:2022awq}. Concerning kinematics, since
the proton mass $M$ explicitly enters the matching expressions for GTMDs on
GPDs, the momentum transfer $t=\Delta^2$ is subject to the following constraint:
\[
|t| \ge |t_{\text{min}}| = \frac{4M^2\xi^2}{1-\xi^2}\,.
\]
Assuming $M\simeq 1$~GeV, as appropriate for protons, in the following we will
show results at $t=-0.5$ GeV$^2$, which in turn implies that $\xi \lesssim
0.35$. Moreover, one needs to specify the angle $\theta_{k\Delta}$ between the
parton transverse momentum ${\bm k}_T$ and the transverse component ${\bm
\Delta}_T$ of the proton momentum transfer (see
App.~\ref{sec_Fourier_transform}). We set it to $\theta_{k\Delta} = 0$ for
$\hat{S}_{1,1a}^{0;+;u}$ and to $\theta_{k\Delta} = \pi/4$ for
$\hat{S}_{1,1b}^{0;+;u}$, which maximises the effect of the mixing between $U$
and $T$ channels.

We point out that the numerical results presented below were obtained using
publicly available codes. Specifically, the GK model for the GPDs at the initial
scale $\mu_0=2$~GeV is provided by {\tt PARTONS}~\cite{Berthou:2015oaw}; their
collinear evolution as well as the perturbative ingredients relevant to matching
and evolution are provided by {\tt APFEL++}~\cite{Bertone:2013vaa,
Bertone:2017gds}; the non-perturbative components as well as the Fourier
transform of GTMDs are provided by {\tt NangaParbat}~\cite{Bacchetta:2019sam}.

\begin{figure}
    \centering
    \includegraphics[width=0.49\linewidth]{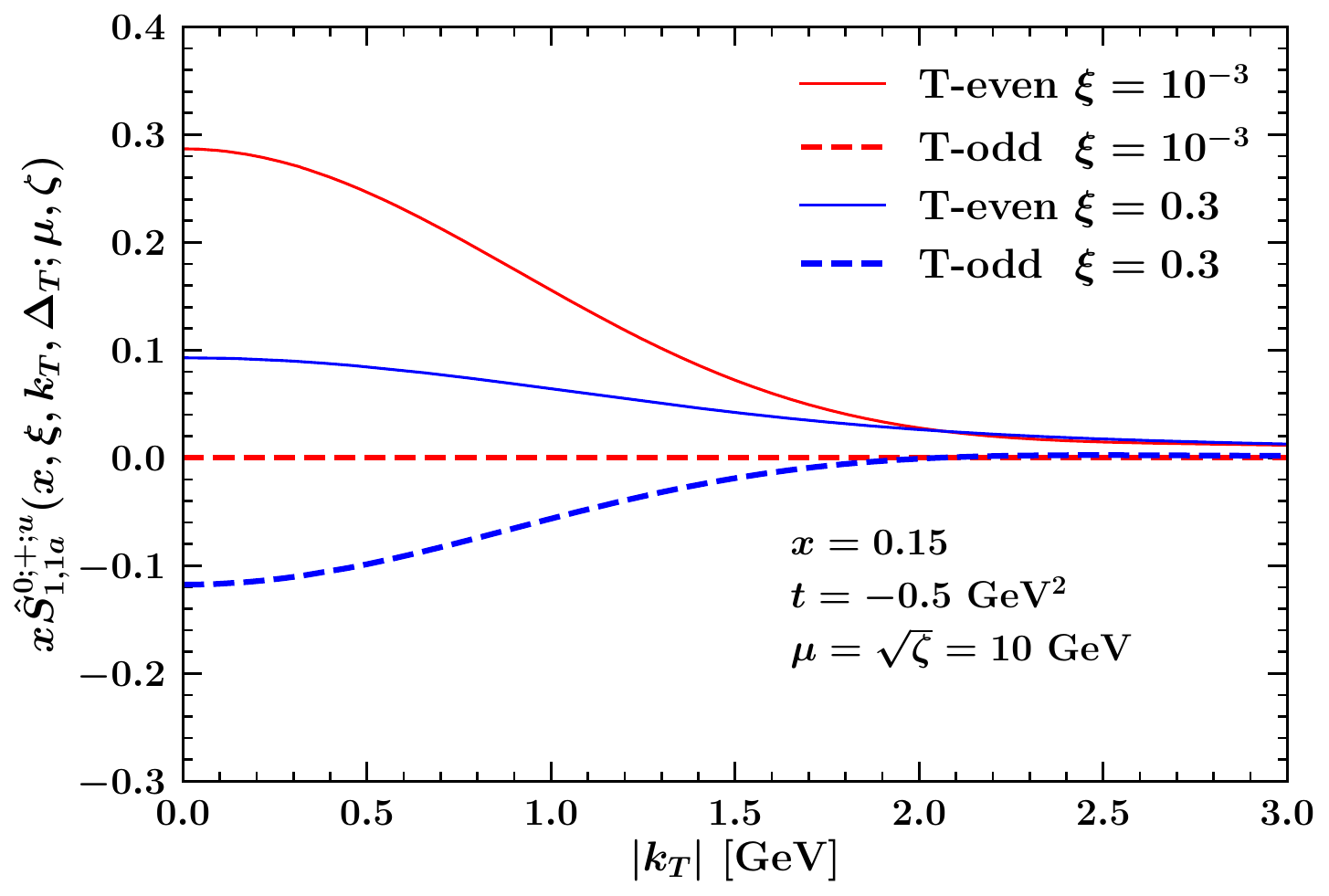}
    \includegraphics[width=0.49\linewidth]{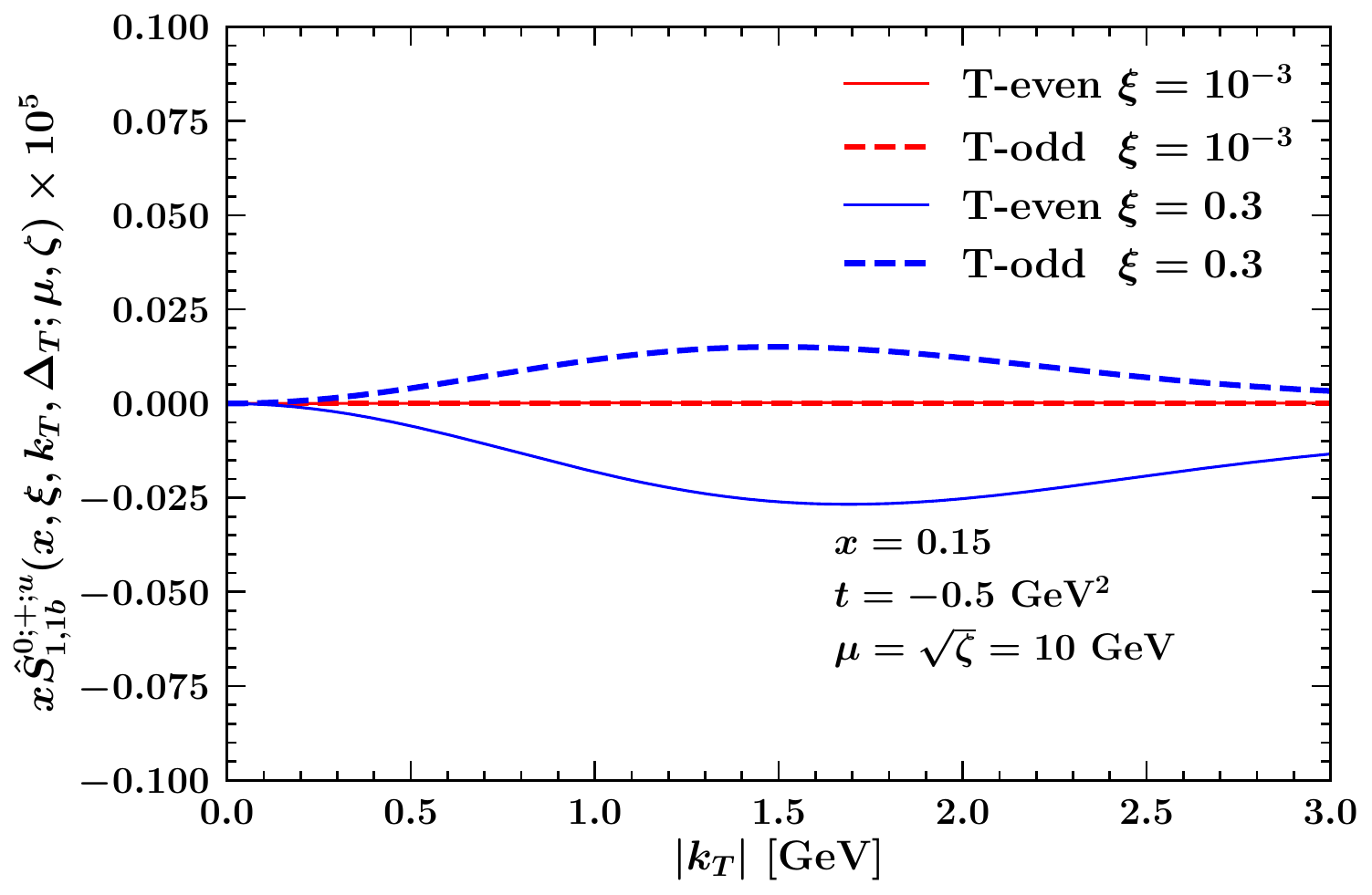}
    \vspace{15pt}
    \caption{The GTMDs $\hat{S}_{1,1a}^{0;+;u}$ (left) and
    $\hat{S}_{1,1b}^{0;+;u}$ (right) plotted as functions of $|{\bm k}_T|$ at
    $x=0.15$ and $\mu=\sqrt{\zeta}=10$~GeV. Two different values of $\xi$ are
    considered: $\xi=10^{-3}<x$ (red curves) which probes the DGLAP region, and
    $\xi=0.3>x$ (blue curves) which instead probes the ERBL region. T-even and
    T-odd components are shown separately as solid and dashed curves,
    respectively. The plot for $\hat{S}_{1,1b}^{0;+;u}$ is magnified by a factor
    of $10^{5}$ for presentation purposes.}
    \label{plot_S11ab0plusUp_x0p2}
\end{figure}

Fig.~\ref{plot_S11ab0plusUp_x0p2} displays the behavior of the up-quark GTMDs
$\hat{S}_{1,1a}^{0;+;u}$ (left plot) and $\hat{S}_{1,1b}^{0;+;u}$ (right plot)
as functions of $|{\bm k}_T|$ at $x=0.15$ and $\mu=\sqrt{\zeta}=10$~GeV for two
different values of $\xi$. These values were chosen such that the first
($\xi=10^{-3} < x$, red curves) corresponds to the DGLAP region, while the
second ($\xi=0.3 > x$, blue curves) probes the ERBL region. T-even and T-odd
contributions to both GTMDs are shown separately as solid and dashed curves,
respectively. The curves for $\hat{S}_{1,1b}^{0;+;u}$ are multiplied by a factor
of $10^{5}$ to make them comparable in size with $\hat{S}_{1,1a}^{0;+;u}$.
First, we note that, consistently with Eq.~(\ref{eq:AfterMatching}), T-odd
contributions are identically zero in the DGLAP regions (see red dashed curves).
In the ERBL region, the T-even component generally dominates for both
distributions, though the T-odd contribution remains non-negligible. This aligns
with the expectation that T-odd terms are suppressed by a power of $\alpha_s$
relative to their T-even counterparts. The suppression of
$\hat{S}_{1,1b}^{0;+;u}$ compared to $\hat{S}_{1,1a}^{0;+;u}$ can be partially
attributed to the same reasoning, along with the factor $c=10^{-1}$, which
suppresses linearly polarised gluon GPDs relative to their unpolarised and
longitudinally polarised counterparts. However, this alone does not account for
the full factor of $10^5$, which is primarily due to the effect of the matching
functions. However, we point out that this suppression does not have any direct
implications for the orbital angular momentum
distribution~\cite{Lorce:2011kd,Hatta:2011ku}, since, as evident from
Eq.~\eqref{eq_S11b0pq_res}, the matching is proportional to ${\bm\Delta}_T$ and
hence does not contribute to the orbital angular momentum distribution (see
Eq.~(28) of Ref.~\cite{Lorce:2011kd}).

\begin{figure}
    \centering
    \includegraphics[width=0.6\linewidth]{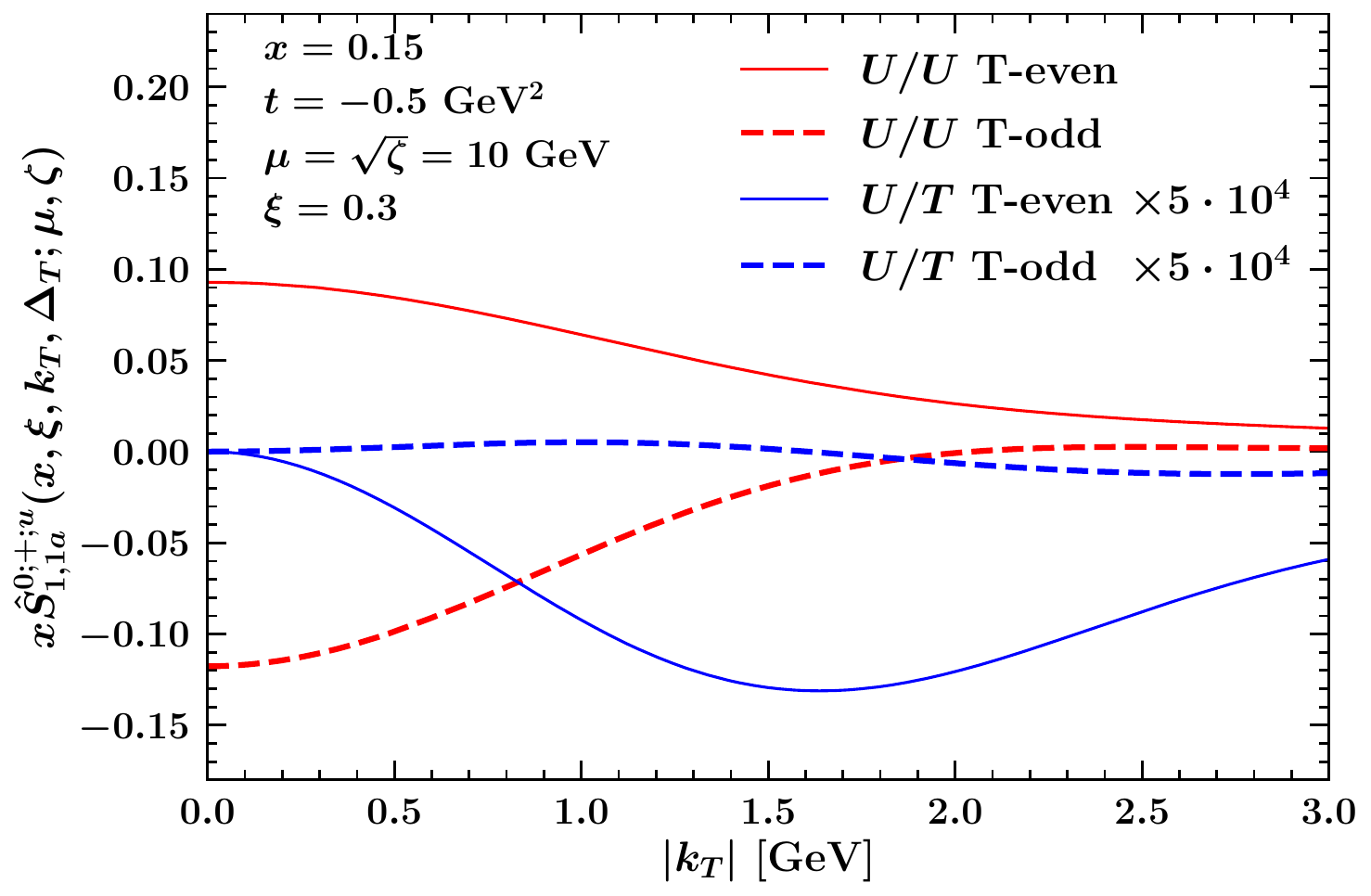}
    \vspace{15pt}
    \caption{The T-even (solid curves) and T-odd (dashed curves) components to
    the GTMD $\hat{S}_{1,1a}^{0;+;u}$ as functions of $|{\bm k}_T|$ at $x=0.15$,
    $\xi=0.3$, and $\mu=\sqrt{\zeta}=10$~GeV. Each component is further
    separated into $U/U$ (red curves) and $U/T$ (blu curves) contributions.}
    \label{plot_S11a0plusUp_x0p2_UU_vs_UT}
\end{figure}

As discussed in Sec.~\ref{sec_matching_op_level}, each GTMD polarisation
receives contributions from different GPDs polarisations. Indeed, from
Eq.~(\ref{S^0+q_11a}) we see that the unpolarised GTMD $\hat{S}_{1,1a}^{0;+;u}$
gets contributions both from unpolarised ($U/U$ channel) and
transversely/linearly polarised ($U/T$ channel) GPDs. It is therefore
interesting to examine their relative importance. To this purpose, in
Fig.~\ref{plot_S11a0plusUp_x0p2_UU_vs_UT} we display $U/U$ (red curves) and
$U/T$ (blue curves) contributions to $\hat{S}_{1,1a}^{0;+;u}$ for both T-even
(solid curves) and T-odd (dashed curves) components. The same settings as in
Fig.~\ref{plot_S11ab0plusUp_x0p2} are used but selecting $\xi=0.3$. Moreover,
the $U/T$ contributions are multiplied by a factor of $5\cdot 10^{4}$ to make
them comparable in size to the $U/U$ contributions. The suppression of the $U/T$
channel relative to the $U/U$ channel can be partially attributed to its
suppression by a power of $\alpha_s$ and that linearly polarised gluon GPDs are
a factor of $c=10^{-1}$ smaller than their unpolarised/longitudinally-polarised
counterparts. However, the suppression observed in
Fig.~\ref{plot_S11a0plusUp_x0p2_UU_vs_UT} is stronger than expected and can
again be traced back to the effect of the matching functions. Additionally, we
find that the T-odd contribution is comparable in magnitude to the T-even one,
emphasizing the need to account for T-odd effects in any future phenomenological
study of GTMDs.

\begin{figure}
    \centering
    \includegraphics[width=0.6\linewidth]{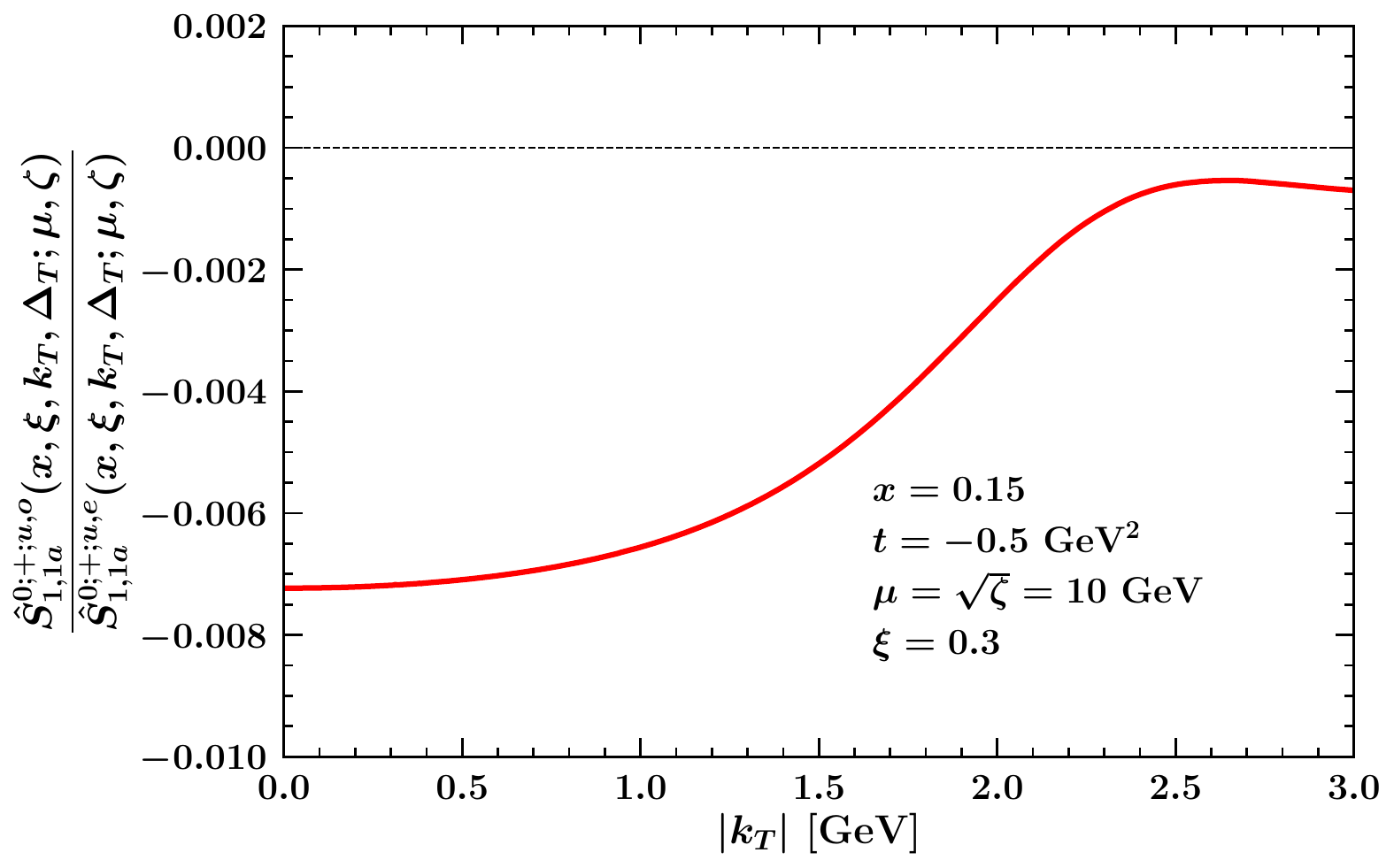}
    \vspace{15pt}
    \caption{Ratio between the T-odd and T-even contributions to
    $\hat{S}_{1,1a}^{0;+;u}$ as a function of $|{\bm k}_T|$ obtained by
    replacing the rotation matrix in Eq.~(\ref{eq:AfterMatching}) with the
    identity matrix. This modification ensures that matching is the sole source
    of T-odd effects.}
    \label{plot_S11a0plusUp_x0p2_only_matching}
\end{figure}

Finally, we investigate the mechanism through which T-odd effects are generated.
As argued above, two different sources of T-odd effects are at play: matching
and evolution. It is therefore insightful to compare their relative
contributions. To this end, Fig.~\ref{plot_S11a0plusUp_x0p2_only_matching}
displays the ratio $\hat{S}_{1,1a}^{0;+;u, o}/\hat{S}_{1,1a}^{0;+;u, e}$ as a
function of $|{\bm k}_T|$, with the rotation matrix in
Eq.(\ref{eq:AfterMatching}) set to unity. This effectively disables the
evolution-induced T-odd effects, isolating those generated by matching. The same
kinematic setup as in Fig.~\ref{plot_S11a0plusUp_x0p2_UU_vs_UT} is used. The
results indicate that the ratio remains close to zero for all values of $|{\bm
k}_T|$ considered, implying that matching-induced T-odd effects contribute less
than 1\% of the total. This proves that evolution is the dominant source of
T-odd effects.

In conclusion, the exploratory numerical study presented above may be useful for
future phenomenological studies in that it exposes various interesting features
of GTMDs. We showed how different GPD polarisations contribute to a given GTMD
polarisation, focussing here on unpolarised GTMDs only. Additionally, we
analysed the interplay between T-even and T-odd components, showing that while
T-odd contributions remain small, they are generally non-negligible in the ERBL
region.  In this respect, we also demonstrated that by far the most important
contribution to the generation of T-odd effects is evolution.

\section{Conclusions}
\label{sec_conclusion}

In this work, we have presented the one-loop, \textit{i.e.}
$\mathcal{O}(\alpha_s)$, corrections to the matching functions for all of the
twist-two GTMDs of the proton. These matching functions are an important
ingredient in any accurate phenomenological study of the internal structure of
the proton. In fact, they allow one to reconstruct GTMDs in terms of GPDs in the
region of small partonic transverse separations ${\bm b}$, or equivalently large
partonic transverse momenta $\mathbf{k}_T$. This is analogous to the matching of
TMDs onto collinear PDFs, which is routinely exploited in extractions of TMDs
from experimental data. In fact, in the largely unexplored realm of GTMDs, their
matching relations on GPDs turn out to be even more important to achieve a
realistic reconstruction of these multidimensional objects.

GTMDs are more complex objects than their unskewed counterparts, \textit{i.e.}
TMDs. Indeed, GTMDs feature a number of novel aspects compared to TMDs, which we
highlighted here for the first time. In Sec.~\ref{sec_matching_op_level}, we
demonstrated that the matching pattern of GTMDs on GPDs is much richer than that
of TMDs. More specifically, we showed that for any given GTMD polarisation
multiple GPD polarisations may contribute (see Tab.~\ref{tab_rij}). We also
demonstrated that one-loop GTMD matching functions on gluon GPDs in the ERBL
region ($\xi>x$) develop an imaginary part proportional to the direction of the
Wilson line, $s=\pm1$, which contributes to the perturbative mixing of T-even
and T-odd GTMDs (see also Eq.~(\ref{eq:InitialScaleMatchingMatrix})). In
Sec.~(\ref{sec_matching_param}), leveraging existing results, we obtained a
parametrisation of the twist-two GTMD correlators in terms of GTMD distributions
in ${\bm b}$ space (suitable for matching) and presented their matching on
standard twist-two GPDs at small $|\bm b|$. In
Sec.~\ref{sec_reconstructingGTMDs}, we focused on the evolution of GTMDs. We
proved that the solution of the evolution equations, more specifically the one
w.r.t. to $\mu$, is responsible for yet another source of T-even/T-odd mixing
(see Eq.~(\ref{eq:BeforeMatching})). Combining perturbative matching and
evolution, we obtained Eq.~(\ref{eq:AfterMatching}), which allows for a full
reconstruction of GTMDs in terms of GPDs and a number of perturbatively
computable quantities. In this formula, a particularly important role is played
by matching functions whose full set of one-loop corrections were computed here
for the first time.

In Sec.~\ref{sec_numerical_results}, we presented a selection of numerical
results with the purpose of quantitatively illustrating the effect of
polarisation as well as that of T-even/T-odd mixing when reconstructing GTMDs.
We found that polarisation mixing plays an important role in the sense that the
contribution of off-diagonal GPDs, \textit{i.e.} GPDs whose polarisation differs
from that of the GTMD they match on, is non-negligible. As far as the
T-even/T-odd mixing is concerned, we found that the mixing produced by evolution
is far larger than the mixing produced by matching.

We finally point out that, modulo the reliability of GPDs and possible
non-perturbative effects, the GTMDs reconstructed in
Sec.~\ref{sec_numerical_results} should be regarded as realistic in that they
encode the full QCD-driven complexity that emerges from perturbative matching
and evolution. Therefore, our results provide a crucial foundation for future
phenomenological studies of GTMDs and their role in understanding the proton
multi-dimensional structure.

\section*{Acknowledgements}

The authors are grateful to M. Diehl for a critical reading of the manuscript.
The work of M.G.E. is supported by the State Research Agency through the grants
PCI2022-132984, PID2022-136510NB-C33 and CNS2022-135186, and by the Basque
Government through the grant IT1628-22. The work of S.R. is supported by the
German Science Foundation (DFG), grant number 409651613 (Research Unit FOR
2926), subproject 430915355.  O. dR. is supported by the MIU (Ministerio de
Universidades, Spain) fellowship FPU20/03110. Additionally, O. dR. thanks DESY
for its hospitality during his three-month research stay, funded by the MIU
mobility grant EST24/00197, during which part of this work was carried out.

\appendix

\section{On the presence of imaginary contributions }\label{AppendixPole}
\newcommand{\sign}[1]{\,\text{sign}\ta {#1}\tc}

\subsection{Imaginary part in the evolution}\label{subsec:twistone}

The imaginary part present in the evolution equation w.r.t. $\mu$ in
Eq.~\eqref{eq:evolutionequationsExp} can be explained in terms of the UV
renormalisation constant for the generic twist-one operator (\textit{e.g.}, see
Eq.~(8.22) of Ref.~\cite{Vladimirov:2021hdn}). In the quark case, its one-loop
expression in $4-2\epsilon$ dimensions reads:
\[
Z\ta \frac{\delta^+}{k^+}\tc = 1+\left(\frac{\alpha_s}{4\pi}\right)\frac{C_F}{\epsilon}\ta\frac{3}{2}+2\log\ta\frac{\delta^+}{isk^+}\tc\tc\,,
\]
where $\delta^+$ is the (dimension-full) regulator of rapidity
divergences\footnote{ The $\delta$-regularisation prescribes that
    the light-like Wilson lines should be modified as follows:
\[
\notag
\text{P}\exp\left[-ign_\mu \int_{s\infty}^a d\sigma A^\mu(n\sigma)\right] \to \text{P}\exp\left[-ign_\mu \int_{s\infty}^a d\sigma A^\mu(n\sigma)e^{- s \delta^+ \sigma}\right]\,,
\]
in which the $|\sigma| \to \infty$ region, responsible for rapidity
divergences, is exponetially suppressed.} (see
Refs. \cite{Echevarria:2011epo,Echevarria:2016scs}), $s=\pm 1$
identifies the direction of the Wilson line, and $k^+$ is the momentum
of the external parton. The UV renormalisation for the whole twist-two
operator associated with the GTMDs is obtained as:
\[
Z^\dag\ta \frac{\delta^+}{k^+_1}\tc Z\ta \frac{\delta^+}{k^+_2}\tc \,,
\]
where $k^+_1$ ($k_2^+$) is the momentum associated with the left (right) field
in Eqs.~\eqref{QuarkCorrelator} and~\eqref{GluonCorrelator}, which we
parametrise as $k_1^+ = (1-\kappa) xp^+$ ($k_2^+ = (1+\kappa) xp^+$). Here, we
stress once again that we can always reduce ourselves to the case of $x,\xi
\ge0$, which implies $\kappa\ge0$. Therefore, the one-loop UV renormalisation
constant takes the form:
\begin{equation}
  \begin{array}{rcl}
    \displaystyle  Z^\dagger\ta \frac{\delta^+}{(1-\kappa)xp^+}\tc Z\ta \frac{\delta^+}{(1+\kappa)xp^+}\tc &=&\displaystyle 1+\left(\frac{\alpha_s}{4\pi}\right)\frac{2C_F}{\epsilon} \Bigg( \frac{3}{2}  + 2 \log\ta\frac{\delta^+}{xp^+}\tc  - \log(|1-\kappa^2|)\\
&-&\displaystyle \log(-is \sign{1-\kappa}) - \log(is \sign{1+\kappa})\Bigg)\,.
  \end{array}
  \label{eq:UVrenTwist2}
\end{equation}
The term $\log(|1-\kappa^2|)$ in the equation above is precisely the same that
appears in the GTMD evolution equation w.r.t. $\mu$ in
Eq.~\eqref{eq:evolutionequationsExp}. The terms in the second line, instead,
generate the imaginary part. Indeed, we have:
\begin{equation}
- \log(-is \sign{1+\kappa}) - \log(is \sign{1-\kappa}) = -i\pi s \theta(\kappa -1)\,,
\end{equation}
such that the imaginary part vanishes in the DGLAP region ($\kappa<1$), while it
survives in the ERBL region ($\kappa>1$).

Now we can complete the renormalization of the GTMD operator by introducing the
renormalization of the rapidity divergences, to obtain:
\[
\mathcal{F}^{[Y]}_{q/H;bare} = R\ta b^2, \frac{\delta^+}{\nu^+}\tc Z^\dagger\ta \frac{\delta^+}{(1-\kappa)xp^+}\tc Z\ta \frac{\delta^+}{(1+\kappa)xp^+}\tc \mathcal{F}^{[Y]}_{q/H}(\nu^+,\mu)\,,
\]
where the scale $\nu^+$ is introduced as a reference scale to subtract rapidity
divergences. The last step is to divide out the (square root of the) soft
factor\footnote{For the sake of illustration, we use a similar notation to Ref. \cite{Vladimirov:2021hdn}, which we refer the reader for further details.}. To this end, one can write the soft factor as (where $\nu^2 = 2\nu^+\nu^-$):
\begin{align}
S(b^2,2\delta^+\delta^-) & = Z_S\ta\frac{2\delta^+\delta^-}{\mu^2}\tc R\ta b^2, \frac{\delta^+}{\nu^+}\tc R\ta b^2, \frac{\delta^-}{\nu^-}\tc S_0(b^2,\nu^2) \\
& \notag = \Bigg[ R\ta b^2, \frac{\delta^+}{\nu^+}\tc \sqrt{Z_S\ta\frac{\nu^2}{\mu^2}\tc Z_R\ta\frac{\delta^+}{\nu^+}\tc S_0(b^2,\nu^2) }\,\,\Bigg]\\
& \notag \times 
\Bigg[ R\ta b^2, \frac{\delta^-}{\nu^-}\tc \sqrt{Z_S\ta\frac{\nu^2}{\mu^2}\tc Z_R\ta\frac{\delta^-}{\nu^-}\tc S_0(b^2,\nu^2) }\,\,\Bigg]\,,
\end{align}
where $S_0$ is free from UV and rapidity divergences and $Z_S$ is the UV
renormalization for the soft factor:
\[
Z_S\ta\frac{2\delta^+\delta^-}{\mu^2}\tc \equiv Z_S\ta\frac{\nu^2}{\mu^2}\tc Z_R\ta\frac{\delta^+}{\nu^+}\tc Z_R\ta\frac{\delta^-}{\nu^-}\tc\,.
\]
Referring to the pion-nucleon double Drell-Yan process detailed in
Ref.~\cite{Echevarria:2022ztg}, the first term in the second line is divided out
from the proton GTMD, the second term in the second line is divided out from the
pion light-cone wave function. Finally, we can write:
\[
\frac{\mathcal{F}^{[Y]}_{q/H;bare}}{R\ta b^2, \frac{\delta^+}{\nu^+}\tc \sqrt{Z_S\ta\frac{\nu^2}{\mu^2}\tc Z_R\ta\frac{\delta^+}{\nu^+}\tc S_0(b^2,\nu^2) }} = \mathbb{Z}\ta \frac{\xi}{x},\frac{\mu^2}{\zeta}\tc \mathcal{F}^{[Y]}_{q/H}(\zeta,\mu)\,,
\]
where:
\[
\mathbb{Z}\ta x,\xi,\frac{\mu^2}{\zeta}\tc = \frac{Z^\dagger\ta \frac{\delta^+}{(1-\kappa)xp^+}\tc Z\ta \frac{\delta^+}{(1+\kappa)xp^+}\tc}{\sqrt{Z_S\ta\frac{\nu^2}{\mu^2}\tc Z_R\ta\frac{\delta^+}{\nu^+}\tc  }}\,,
\]
and:
\[
\mathcal{F}^{[Y]}_{q/H}(\zeta,\mu) = \frac{\mathcal{F}^{[Y]}_{q/H}(\nu^+,\mu)}{\sqrt{S_0(b^2,\nu^2)}}\,.
\]
In the relations above, we introduced the rapidity scale $\zeta$ as:
\[
\zeta = 2 |k_1^+ k_2^+| \frac{\nu^-}{\nu^+} = 2 (xp^+)^2|1-\kappa^2|\frac{\nu^-}{\nu^+}\,.
\]
This scale cannot be fixed completely independently from the other components of
the factorisation theorem, since it must satisfy the equality:
\[
\zeta \bar{\zeta} = (Q_1^2)(Q_2^2)\,,
\]
where $Q_{1,2}^2$ are hard scales equal to the momenta of the two virtual bosons
and $\bar{\zeta}$ has a similar definition for the light-cone wave function of
the pion.

As a final remark, we stress that the emergence of the imaginary part in the UV
renormalization factors is \textit{not} an artifact of the specific regulator
$\delta^+$. Regulating rapidity divergences using off-the-light-cone Wilson
lines leads to the same result.\footnote{Compare with the UV pole part of the
one-loop coefficient function for quasi-TMD operators in Eq.~(4.31) of
Ref.~\cite{Rodini:2022wki}.} For a more in-depth discussion about the
intricacies of the definition of the rapidity scale, we refer to
Refs.~\cite{Buffing:2017mqm,Diehl:2021wpp}. 

\subsection{Imaginary part in the residual function}

We now discuss the terms $is\delta$ in the results of Tab. \ref{tab_rij}, which
in turn give rise to an imaginary part in the matching functions through
Eq.~(\ref{eq:MatchingDecompositionforSinglet}). In order to sketch how this
imaginary part emerges, we first note that it is only present in matching
functions that are to be convoluted with a gluon GPD. In the parton-in-parton
approach and at leading twist, these matching functions can be obtained by
computing matrix elements between \textit{physical} gluon states. The
computation of any such matching function necessarily boils down to evaluating
matrix elements of this sort:
\[
\phantom{}_g\braket{k_1|A^j(x_1)A^i(x_2)|k_2}_g
\label{eq_Cxg_matrix_element}\,,
\]
where the subscripts $g$ indicate gluon states. To relate this matrix element to
the gluon GPD in Eq.~\eqref{GluonGPDCorrelator}, we need to express it in terms
of the field strength tensor $F^{+i}$. The simplest way to present the argument
is to work in light-cone gauge $A^+=0$, equipped with appropriate boundary
conditions:
\[
\lim_{L\to s\infty}A^i(x+Ln) = 0\,.
\]
Importantly, $s$ has to be the same sign that determines the direction of the
Wilson line in the GTMD correlator. Indeed, in order to reduce to unity the
transverse gauge link at light-cone infinity involved in the GTMD correlator,
$A^i(x+Ln)$ must vanish exactly when $L \rightarrow s\infty$. This allows us to
write:
\[
A^i(x) = \lim_{L\to s\infty}  \int^0_L d\sigma F^{+i}(x+\sigma n) 
\label{eq_from_A_to_F}\,.
\]
With a slight abuse of notation, Eq.~\eqref{eq_from_A_to_F} in momentum space
becomes:
\begin{align}
A^i(k) e^{-i(kx)} &= F^{+i}(k)  e^{-i(kx)}\int^0_{s\infty} d\sigma e^{-i(k^+-is\delta p^+)\sigma}\,,
\end{align}
where we performed the shift:\footnote{We dropped the shift in $e^{-i(kx)}$,
since it is uninfluential for the argument and thus can be safely discarded.}
\[
k^+ \to k^+ -is\delta p^+\,,
\]
with $\delta$ a small positive parameter. Evaluating the integral in $\sigma$,
we obtain:
\[
A^i(k) = \frac{i}{k^+-i\delta sp^+} F^{+i}(k)
\label{how_to_get_F_from_A_0}\,.
\]
Using this identity in the momentum-space version of
Eq.~(\ref{eq_Cxg_matrix_element}), we find:
\[
\phantom{}_g\braket{k_1|A^j(k_1)A^i(k_2)|k_2}_g=\frac{1}{k_2^+-is\delta p^+}\frac{1}{k_1^++is\delta p^+} \phantom{}_g\braket{k_1|F^{+j}(k_1)F^{+i}(k_2)|k_2}_g 
\label{how_to_get_F_from_A}
\,,
\]
which achieves the goal of rewriting the matrix element in
Eq.~(\ref{eq_Cxg_matrix_element}) in a form compatible with the definition of
gluon GPD in Eq.~\eqref{GluonGPDCorrelator}. As customary, the gluon momenta
$k_1$ and $k_2$ can be chosen to have only non-vanishing $+$ components, which
are parametrised as $k_{1}^+=(1-\eta)(x/y)p^+$ and $k_2^+=(1+\eta)(x/y)p^+$. The
variable $\eta$ denotes the ratio between skewness $\xi$ and GPD partonic
longitudinal momentum. This is a different variable from $\kappa=\xi/x$. Indeed,
using Eq.~\eqref{MatchingOPE} as a reference, one has that $\eta = \xi / (x/y) =
\kappa y$. The consequence is that the matrix element in the l.h.s. of
Eq.~\eqref{how_to_get_F_from_A} will be proportional to the factor: 
\[
  \frac{1}{1-\kappa y+is\delta}\frac{1}{1+\kappa y-is\delta}\,,
  \label{eq:MatrixElementFactor}
\]
which is singular at $y=\pm 1/\kappa$. Note that we have extracted a factor
$(x/y)p^+$ from each denominator without having to keep track of its sign in the
imaginary part, since it is always possible to choose $x, y, \kappa>0$. This has
also the consequence of leaving only the singularity at $y=1/\kappa$ inside the
integration range, while pushing the singularity at $y=-1/\kappa$ outside of it.
Therefore, we can remove the regulator $\delta$ from the second factor in
Eq.~(\ref{eq:MatrixElementFactor}) and treat it as a regular function. Finally,
using the identity:
\begin{equation}
  \frac{1}{1-\kappa y+is\delta} = \text{PV}\ta\frac{1}{1-\kappa y}\tc - is\frac{\pi}{\kappa} \delta\ta
  y-\frac{1}{\kappa}\tc\,,
  \label{eq:PVdefinition}
\end{equation}
valid for $\delta$ vanishingly small, we obtain the imaginary part in the
matching functions in Eq.~(\ref{eq:MatchingDecompositionforSinglet}), which,
crucially, is proportional to the Wilson-line direction $s$.

A similar argument also applies to the calculation of the one-loop GPD splitting
functions $\mathcal{P}_{i/j}^{\Gamma,[0]}$~\cite{Bertone:2022frx,
Bertone:2023jeh} which appear in Eq.~\eqref{DecompositionMatchingCoefficient}.
However, it turns out that in this case the coefficient of $(1-\kappa
y+is\delta)^{-1}$ vanishes, implying the absence of imaginary contributions
related to the computation of matrix elements as in
Eq.~(\ref{eq_Cxg_matrix_element}).

\subsection{Vanishing imaginary part in the splitting functions}

There is yet another possible source of imaginary contributions that may affect
the one-loop splitting functions $\mathcal{P}_{i/j}^{\Gamma,[0]}$, namely the
imaginary terms that emerge from the renormalisation of the twist-one operators
discussed in Sec.~\ref{subsec:twistone} (see Eq.~(\ref{eq:UVrenTwist2})). We now
review the proof that splitting functions are free of rapidity divergences for
$\delta \rightarrow 0$. In the proof, we keep explicit track of the various
imaginary parts, showing that they eventually cancel out leaving a real result.
To do so, we limit to the unpolarised non-singlet splitting function
$\mathcal{P}^{U,-,[0]}$ (all the others follow a similar pattern) and, without
loss of generality, we assume a future-pointing Wilson line ($s=1$). The
starting point reads~\cite{Bertone:2022frx, Bertone:2023jeh}:
\begin{equation}
 \mathcal{P}^{U,-,[0]}\left(y,\kappa\right) = \theta(1-y)
 \mathcal{P}_1^{U,-,[0]}\left(y,\kappa\right)+\theta(\kappa-1)
 \mathcal{P}_2^{U,-,[0]}\left(y,\kappa\right)\,,
\label{eq:SplittingDecompositionforNonSinglet}
\end{equation}
where:
\begin{equation}
\begin{array}{rcl}
  \displaystyle\mathcal{P}_1^{U,-,[0]}\left(y,\kappa\right)&=&\displaystyle
                                                                    p^U_{q/q}\left(y,\kappa\right)+p^U_{q/q}\left(y,-\kappa\right)\\
&+&\displaystyle \vphantom{\Bigg|}\delta(1-y) 2C_q\left[\frac{3}{2}+\log\left(\frac{-i\delta}{1+\kappa}\right)+\log\left(\frac{-i\delta}{1-\kappa}\right)\right]\,,\\
\mathcal{P}_2^{U,-,[0]}\left(y,\kappa\right)&=&\displaystyle
                                                     -p^U_{q/q}\left(y,-\kappa\right)
                                                     +p^U_{q/q}\left(-y,-\kappa\right)\,.
\label{eq:P1andP2NS}
\end{array}
\end{equation}
We can read off $p^U_{q/q}$ from Ref.~\cite{Bertone:2022frx}:
\begin{equation}
  p^U_{q/q}(y,\kappa) = C_F \frac{(1+\kappa)
                       (1-y+2\kappa y)}{\kappa (1+\kappa y)(1-y-i\delta)}\,.
\label{eq:explicitpqq}
\end{equation}
Since $\mathcal{P}_1^{U,-,[0]}$ in
Eq.~(\ref{eq:SplittingDecompositionforNonSinglet}) is accompanied by
$\theta(1-y)$, the singularity of $p^U_{q/q}$ at $y=1$ is an end-point one. In
this case, we use the identity:
\begin{equation}
  \frac{1}{1-y- i\delta}\rightarrow
  \left(\frac{1}{1-y}\right)_+-\delta(1-y) \log\left(-i\delta\right)\,,
\label{eq:cdeltaregularisation}
\end{equation}
where the $+$-prescription is defined as in Eq.~(40) of
Ref.~\cite{Bertone:2022frx}, so that:
\begin{equation}
  p^U_{q/q}(y,\kappa) = C_F \frac{(1+\kappa)
                       (1-y+2\kappa y)}{\kappa (1+\kappa y)}\left(\frac{1}{1-y}\right)_+-\delta(1-y)2C_F\log(-i\delta)\,.
\end{equation}
When inserted into the first equation in Eq.~(\ref{eq:P1andP2NS}), this
produces:
\begin{equation}
\displaystyle \mathcal{P}_1^{U,-,[0]}\left(y,\kappa\right) =  p^U_{q/q}(y,\kappa)+p^U_{q/q}(y,-\kappa) + \delta(1-y)
    2C_F\left[\frac{3}{2}-\log\left(1+\kappa\right)-\log\left(1-\kappa\right)\right]\,,
    \label{eq:P1Um0}
\end{equation}
where $p^U_{q/q}$ here is understood to have the factor $(1-y)^{-1}$ replaced by
$(1-y)_+^{-1}$. This leaves us with a term proportional to
$\log\left(1-\kappa\right)$ that, for $\kappa>1$ (\textit{i.e.} in the ERBL
region), becomes complex.

If we now consider $\mathcal{P}_2^{U,-,[0]}$, the singularity at $y=1$ is no
longer an end-point one. This allows us to replace the factor
$(1-y-i\delta)^{-1}$ in Eq.~(\ref{eq:explicitpqq}) using
Eq.~(\ref{eq:PVdefinition}) with $\kappa=1$, so that:
\begin{equation}
  p^U_{q/q}(y,-\kappa) =- C_F \frac{(1-\kappa)
    (1-y-2\kappa y)}{\kappa (1-\kappa y)}\mbox{PV}\left(\frac{1}{1-y}\right)+i\pi\delta(1-y) 2C_F \,,
\end{equation}
which gives:
\begin{equation}
\mathcal{P}_2^{U,-,[0]}\left(y,\kappa\right)=- p^U_{q/q}(y,-\kappa)+p^U_{q/q}\left(-y,-\kappa\right)-i\pi\delta(1-y) 2C_F\,,
\label{eq:P2Um0}
\end{equation}
where this time $p^U_{q/q}(y,-\kappa)$ has the factor $(1-y)^{-1}$ replaced by
$\mbox{PV}(1-y)^{-1}$.

Now, if we take the combination in
Eq.~(\ref{eq:SplittingDecompositionforNonSinglet}) using Eqs.~(\ref{eq:P1Um0})
and~(\ref{eq:P2Um0}), we see that the $\delta$-function terms combine as
follows:
\begin{equation}
2C_F\left[\frac{3}{2}-\log\left(1+\kappa\right)-\log\left(1-\kappa\right)-i\pi\theta(\kappa-1)\right]=2C_F\left[\frac{3}{2}-\log\left(|1-\kappa^2|\right)\right]\,,
\end{equation}
which finally proves that the splitting function $\mathcal{P}^{U,-,[0]}$ is free
of imaginary contributions, \textit{i.e.} it is real.

\section{Fourier transformation}
\label{sec_Fourier_transform}

In this section, we work out explicitly two examples of Fourier transforms
necessary to convert GTMD distributions from $\bm{b}$-space to $\bm{k}_T$-space.
We consider the GTMDs $S_{1,1a}^{0;+;i}$ and $S_{1,1b}^{0;+;i}$ defined in
Eq.~\eqref{eq:ParemtrizationUnpolGTMD} and showed in
Sec.~\ref{sec_numerical_results} for $i=q$. In the first case, we need to
compute the integral:
\[
\hat{S}_{1,1a}^{0;+;i}(x,\xi,\bm{k}_T, \bm{\Delta}_T;\mu,\zeta) = \int d^2\bm{b}\,e^{-i\bm{b}\cdot\bm{k}_T} S_{1,1a}^{0;+;i}(x,\xi,\bm{b},\bm{\Delta}_T;\mu,\zeta)\,.
\label{eq:FourierTransformApp2}
\]
A $\bm{b}$-dependent structure comes from the last line of
Eqs.~\eqref{S^0+q_11a} and \eqref{S^0+g_11a}. If we combine these equations with
Eq.~\eqref{eq:GTMDevolutionSolved}, which encodes the evolution of GTMDs, we can
write:
\begin{equation}
    \begin{array}{rcl}
    S_{1,1a}^{0;+;i}(x,\xi,\bm{b},\bm{\Delta}_T;\mu,\zeta) &=& \displaystyle \mathcal{R}_i\left[(\mu,\zeta)\leftarrow
    (\mu_{b},\mu_{b}^2)\right]f_1^i(x,\xi,\bm{\Delta}_T;\mu_b)\\
    &-&\displaystyle\vphantom{\Bigg|}\mathcal{R}_i\left[(\mu,\zeta)\leftarrow
    (\mu_{b},\mu_{b}^2)\right]\frac{2(\bm{b}\cdot\bm{\Delta}_T)^2-\bm{b}^2\bm{\Delta}_T^2}{4\bm{b}^2M^2} f_2^i(x,\xi,\bm{\Delta}_T;\mu_b)\,,
    \end{array}
\end{equation}
where $f_1$ and $f_2$ can be read off from Eqs.~\eqref{S^0+q_11a} and
\eqref{S^0+g_11a} and are given by:
\begin{align}
\notag f_1^i (x,\xi,\bm{\Delta}_T;\mu_b)=&\;\mathcal{C}_{i/q}^{U/U}\otimes \qa (1-\xi^2)H^q -\xi^2E^q\qc(x,\xi,\bm{\Delta}_T;\mu_b)\\
&+\mathcal{C}_{i/g}^{U/U}\otimes \qa (1-\xi^2)H^{g} -\xi^2E^{g}\qc(x,\xi,\bm{\Delta}_T;\mu_b)\,,\\
\notag f_2^i (x,\xi,\bm{\Delta}_T;\mu_b) =&\;\mathcal{C}_{i/g}^{U/T}\otimes\qa E_T^g -\xi \widetilde{E}^g_T+2\widetilde{H}^g_T\qc(x,\xi,\bm{\Delta}_T;\mu_b)\,.
\end{align}

It is generally not possible to compute the Fourier transform in
Eq.~(\ref{eq:FourierTransformApp2}) fully analytically. However, it can be
reduced to a Hankel transform, which can be evaluated numerically. This leads
to:
\begin{align}
\notag&\hat{S}_{1,1a}^{0;+;i}(x,\xi,\bm{k}_T,\bm{\Delta}_T;\mu,\zeta)=2\pi\int_0^\infty d|\bm{b}|\,|\bm{b}|J_0(|\bm{b}||\bm{k}_T|)\mathcal{R}_i\left[(\mu,\zeta)\leftarrow
    (\mu_{b},\mu_{b}^2)\right]f_1^i(x,\xi,\bm{\Delta}_T;\mu_b)\\
&+ 2\pi\,\frac{2(\bm{k}_T\cdot\bm{\Delta}_T)^2-\bm{k}_T^2\bm{\Delta}_T^2}{4M^2\bm{k}_T^2}\int_0^\infty d|\bm{b}|\,|\bm{b}|J_2(|\bm{b}||\bm{k}_T|)\mathcal{R}_i\left[(\mu,\zeta)\leftarrow
    (\mu_{b},\mu_{b}^2)\right]f_2^i(x,\xi,\bm{\Delta}_T;\mu_b)\,.
\end{align}
The prefactor of the term in second line of the equation above can be rewritten
as follows:
\[
 2\pi\,\frac{2(\bm{k}_T\cdot\bm{\Delta}_T)^2-\bm{k}_T^2\bm{\Delta}_T^2}{4M^2\bm{k}_T^2} = 2\pi  \left[ (1-\xi^2)\frac{(-t)}{4M^2} - \xi^2\right]\cos(2\theta_{k\Delta}),
\]
being $t=\Delta^2<0$ and $\theta_{k\Delta}$ the angle between $\bm{k}_T$ and
$\bm{\Delta}_T$.

In the case of $S_{1,1b}^{0;+;i}$, the relevant term to transform is:
\[
\hat{S}_{1,1b}^{0;+;i}(x,\xi,\bm{k}_T, \bm{\Delta}_T;\mu,\zeta) \equiv \int d^2\bm{b}\,e^{-i(\bm{b}\cdot\bm{k}_T)}\frac{i\e_T^{b\Delta_T}}{M|\bm{b}|}S_{1,1b}^{0;+;i}(x,\xi,\bm{b},\bm{\Delta}_T;\mu,\zeta)\,.
\]
The GTMDs in $\bm{b}$-space from Eqs.~\eqref{eq_S11b0pq_res} and
\eqref{eq_S11b0pg_res} are given by:
\[
S_{1,1b}^{0;+;i}(x,\xi,\bm{b},\bm{\Delta}_T;\mu,\zeta) = \mathcal{R}_i\left[(\mu,\zeta)\leftarrow
    (\mu_{b},\mu_{b}^2)\right]\frac{(\bm{b}\cdot\bm{\Delta}_T)}{2M|\bm{b}|} f_3^i(x,\xi,\bm{\Delta}_T;\mu_b)\,,
\]
where $f_3$ reads:
\begin{align}
f_3^i (x,\xi,\bm{\Delta}_T;\mu_b) =&\mathcal{C}_{i/g}^{U/T}\otimes\qa \xi E_T^g - \widetilde{E}^g_T\qc(x,\xi,\bm{\Delta}_T;\mu_b)\,.
\end{align} 
Therefore, its Fourier transform yields:
\begin{align}
\notag\hat{S}_{1,1b}^{0;+;i}(x,\xi,\bm{k}_T, \bm{\Delta}_T;\mu,\zeta)=&-\frac{\pi}{M^2}\,\frac{i\e_T^{k_T\Delta_T}(\bm{k}_T\cdot \bm{\Delta}_T)}{\bm{k}_T^2}\\
&\int_0^\infty d|\bm{b}|\,|\bm{b}|J_2(|\bm{b}||\bm{k}_T|)\mathcal{R}_i\left[(\mu,\zeta)\leftarrow
    (\mu_{b},\mu_{b}^2)\right]f_3^i(x,\xi,\bm{\Delta}_T;\mu_b)\,,
\end{align}
where the prefactor can be conveniently expressed as follows:
\[
-\frac{\pi}{M^2}\,\frac{i\e_T^{k_T\Delta_T}(\bm{k}_T\cdot \bm{\Delta}_T)}{\bm{k}_T^2} = -2\pi i \left[ (1-\xi^2)\frac{(-t)}{4M^2} - \xi^2\right]\sin(2\theta_{k\Delta}).
\]

\bibliographystyle{ieeetr}

\end{document}